\documentclass[]{jfm}
\usepackage{color}

\usepackage{graphicx}
\usepackage{newtxtext}
\usepackage{newtxmath}
\usepackage{natbib}
\usepackage{hyperref}
\hypersetup{
    colorlinks = true,
    urlcolor   = blue,
    citecolor  = black,
}

\newcommand{\RomanNumeralCaps}[1]

\title{Surrogate Modeling of Urban Boundary-Layer Flows}

\author{Gurpreet S.~Hora\aff{1} and 
  Marco G.~Giometto\aff{1}
   \corresp{\email{mg3929@columbia.edu}}}

\affiliation{\aff{1} Department of Civil Engineering and Engineering Mechanics, Columbia University, New York, NY 10027, USA}

\begin{document}
\maketitle
\begin{abstract}
Surrogate modeling is a viable solution for applications involving repetitive evaluations of expensive computational fluid dynamics models, such as uncertainty quantification and inverse problems.
This study proposes a multi-layer perceptron (MLP) based machine-learning surrogate for canopy flow statistics accommodating any approaching mean-wind angle. 
The training and testing of the surrogate model is based on results from large-eddy simulations of open-channel flow over and within surface-mounted cubes under neutral ambient stratification. 
The training dataset comprises flow statistics from various approaching mean-wind angles, and the surrogate is asked to ``connect between the dots,'' i.e., to predict flow statistics for unseen values of the approaching mean-wind angle.
The MLP performance is compared against a more traditional spline-based interpolation approach for a range of training data.
In terms of relative mean absolute errors on individual flow statistics, the proposed MLP surrogate consistently outperforms the spline interpolation, especially when the number of training samples is reduced. 
The MLP model accurately captures mean profiles and three-dimensional flow variability, offering robust predictions, even when trained with as few as four approaching wind angles.
The model is $10^4 \times$ faster than large-eddy simulations, thus proving effective for multi-query tasks in the context of urban canopy flow modeling. 

\end{abstract}


\section{Introduction}
\label{sec:intro}
The physical structure of cities controls and modifies the exchange of momentum, heat, water, and air pollutants between the land surface and the atmosphere \citep{belcher2005mixing}.
Accurately predicting these exchanges is crucial across a wide range of applications, including local weather forecasting \citep{Shamarock2008}, climate projections \citep{ murakami1999cfd, mochida2008prediction, toparlar2015cfd}, air-quality monitoring \citep{lee1997computational, vardoulakis2003modelling, li2006recent, boppana2010large}, and urban climate studies \citep{chen2012research, krayenhoff2020multi}, to name but a few. 
The intricate interplay between turbulent airflow and urban geometry governs these complex processes.
This interaction has been the subject of extensive research in recent decades, encompassing computational \citep{bou2009effects, li2016quality, li2019contrasts, omidvar2020plume, auvinen2020study, cheng2021turbulent}, experimental \citep{raupach1980wind, rotach1994determination, brown2001comparison, castro2006turbulence, gromke2008dispersion, pascheke2008wind}, and observational \citep{rotach1993turbulence, rotach1999influence, kastner2004mean, rotach2005bubble, christen2009budget, gubler2021evaluation} approaches. 

From a computational perspective, large-eddy simulation (LES) has become the de-facto standard for characterizing microscale turbulent transport in urban areas \citep{xie2006and, kanda2006large, xie2009large, maronga2014effect, giometto2016spatial}. 
LES studies of flow over urban-like topographies have provided insights into the dependence of flow statistics on urban canopy geometry \citep{bou2009effects,yang2016exponential}, characterized the topology of coherent atmospheric structures \citep{kanda2004large, li2024structure}, and elucidated mechanisms responsible for scalar transfer in urban areas \citep{cheng2016large, li2019contrasts}.
Recent years have also seen LES algorithms becoming more readily available and capturing a large range of physical processes \citep[see, for example, ][]{heus2010formulation, maronga2015parallelized, van2017microhh, schmid2024boundarylayerdynamics}. 
With the ever-increasing availability of spatially distributed data \citep{kumar2015rise, middel2022urban}, researchers are presented with a unique opportunity to combine measurements and LES to address open challenges in the aforementioned fields. 
Data assimilation and uncertainty quantification techniques are essential tools to enable such a task \citep{ fletcher2022data}.
However, these multi-query techniques typically require a large number of simulations \citep{smith2013uncertainty}, making LES-based approaches prohibitively expensive \citep{choi2012grid}.

To overcome this challenge, cost-effective approximations of the LES solution are necessary. 
Surrogate models are designed specifically for this specific purpose. 
Surrogate models are simplified representations of complex physical systems that can serve as computationally efficient alternatives to LES.
According to \cite{razavi2012review}, there are two types of surrogates: response surface modeling and lower-fidelity modeling.
Response surface surrogates, known as metamodels \citep{blanning1975construction, kleijnen2009kriging}, model emulation \citep{o2006bayesian}, and proxy models \citep{bieker2007real} utilize data-driven techniques to approximate either the entire computational model or just a part of it. 
Examples of these models include techniques such as Kriging \citep{webster2007geostatistics}, polynomial chaos expansion \citep{wiener1938homogeneous}, and neural networks \citep{goodfellow2016deep}.
These approaches are typically non-intrusive and leverage collected data to effectively model the input-output relationship of a complex physical system, circumventing the need for explicit reliance on the system's underlying governing equations.
Successful studies involved response surface surrogates include \cite{bau2006stochastic, zhu2019machine, enderle2020non, maulik2021turbulent}.
An alternative to using a non-intrusive surrogate model is to use lower-fidelity surrogates. 
These approaches rely on order reduction techniques that are grounded in physics and mathematics to enable an efficient evaluation of the system response \citep{razavi2012review, garzon2022machine}. 
Lower-fidelity surrogates are intrusive and typically require in-depth modifications of the physical solver, but retain reasonable physical characteristics and may be better suited for extrapolation tasks, such as emulating unseen regions of the parameter space \citep{razavi2012review, garzon2022machine}.
Based on this definition, approaches such as Reynolds-averaged Navier-Stokes, proper orthogonal decomposition, and dynamic mode decomposition techniques can be understood as low-fidelity surrogates of direct numerical simulations \citep{Sagaut2006, wilcox1998turbulence, schmid2010dynamic, berkooz1993proper}.

Recently, machine learning (ML), a subset of artificial intelligence, has emerged as a promising approach for solving forward and inverse problems in CFD.
Forward problems in CFD consist of evaluating the solution of partial differential equations with known parameters, initial, and boundary conditions  \citep{bar2019learning, kochkov2021machine, jeon2022finite}. 
Inverse problems, on the other hand, involve inferring quantities of interest, such as initial and boundary conditions or unknown parameters characterizing the physical system under consideration, based on observations of the system \citep{liu2020deep, fukami2021machine, kim2021unsupervised, yousif2023deep}.
ML has also spurred advances in the context of surrogate modeling due to its capability to capture non-linear relationships between inputs and outputs \citep{zhu2019physics, zhu2019machine, ganti2020data, tang2020deep, palar2021gaussian, maulik2021turbulent, nikolopoulos2022non}.
These models are typically supervised using a discrete subset of CFD output data, and once trained and validated, they can be used to evaluate the system response efficiently. 
For instance, \cite{maulik2021turbulent} developed a non-intrusive surrogate model based on a multi-layer perceptron (MLP) architecture to predict the eddy viscosity field in the context of Reynolds-averaged Navier-Stokes equations.
They showed that surrogate models offer a viable approach to exploring vast parameter spaces efficiently.
\cite{zhu2019physics} proposed physics-constrained convolutional neural networks for surrogate modeling of partial differential equation systems. 
They demonstrated that the proposed framework could be trained without labeled or training data by integrating fundamental physics laws and domain knowledge through constraint learning \citep{stewart2017label}.
In general, fusing physics and data enables models to make reliable predictions under both interpolative (training and test inputs are from the same distribution) and extrapolative (test input is out-of-distribution) conditions \citep{karniadakis2021physics}. 

ML has also been recently employed to study turbulent transport in urban environments. 
Researchers have developed non-intrusive reduced-order models using ML to analyze three-dimensional instantaneous flow fields within urban environments \citep{xiao2019reduced, xiang2021fast}. 
Generative adversarial networks, a type of ML technique, have been used to enhance the resolution (super-resolution) of atmospheric quantities such as wind, solar radiation, and air temperature \citep{stengel2020adversarial, wu2021deep}.
Further, several studies have explored using ML to create predictive models that estimate climate variables from urban morphology \citep{javanroodi2022combining, lu2023using}.
For instance, \cite{lu2023using} developed an encoder-decoder convolutional neural network to predict the time-averaged streamwise mean velocity field from arbitrary urban geometries at a fixed approaching wind angle.
Findings demonstrated that ML can accurately predict the spatial variability of the streamwise mean velocity field for unseen idealized urban geometry.

Building from these studies, this work proposes a MLP surrogate for the prediction of time-averaged, three-dimensional flow statistics in an idealized urban canopy as a function of the approaching wind angle, under neutrally-stratified flow conditions.
In real-world environments, the approaching wind angle is continuously changing, and these variations significantly impact spatially distributed flow statistics. 
The geometry of the canopy consists of an array of aligned cubes with uniform height and packing density. 
Such a surface entails key modes of variability of urban environments (presence of flow separation, flow canyoning, wake flow, etc.) while still remaining amenable for comprehensive analysis. 
This controlled setup provides an optimal benchmark for testing the proposed MLP surrogate and meaningfully discussing its strengths and weaknesses before delving into more complicated tasks (such as, e.g., predicting flow statistics for arbitrary canopy morphologies or ambient stratifications).
Model performance is compared against predictions from a more conventional method based on spline (SPL) interpolation.

The proposed surrogate is trained and evaluated using an extensive high-fidelity LES dataset of flow over the considered urban environment featuring a range of approaching wind angles. 
Three training datasets are considered for the training of the model, each comprising an increasing number of data (small-, moderate-, and big-data regimes). 
The performance of the model is then evaluated against unseen data (test dataset) using standard ML metrics, such as relative mean absolute error (RMAE), as well as turbulent flow statistics.

This paper is structured as follows. 
The numerical algorithm and dataset are described in \S \ref{sec:numerical_setup} and \ref{sec:high-fidelity-dataset}, respectively. 
\S \ref{sec:surrogate-modeling} describes the MLP and SPL surrogates. 
Model predictions are examined in \S \ref{sec:results} and further discussed in \S \ref{sec:discussion}. 
Concluding remarks are drawn in \S \ref{sec:conclusions}.

\section{Methodology}
\label{sec:methods} 

\subsection{Numerical Setup}
\label{sec:numerical_setup}

The filtered Navier-Stokes equations for incompressible and Newtonian fluids are solved in their rotational form \citep{orszag1975numerical} to ensure the conservation of energy in the inviscid limit, i.e.,
\begin{equation}
\begin{cases}
   \frac{\partial u_i}{\partial t} 
      + u_j ( \frac{\partial u_i}{\partial x_j} 
      - \frac{\partial u_j}{\partial x_i} ) 
      = 
      - \frac{\partial \pi}{\partial x_i} 
      - \frac{\partial \tau_{ij}^{SGS}}{\partial x_j} 
      - \Pi_i  + f_i^{\Gamma_{\mathrm{b}}} & \text{in $\Omega \times [0,T]$} \, , \\
   	\frac{\partial u_i}{\partial x_i} =0 & \text{in $\Omega \times [0,T]$} \, , \\
	\frac{\partial u}{\partial z} = \frac{\partial v}{\partial z} = 
	w = 0 & \text{in $\Gamma_{\mathrm{t}} \times [0,T]$} \, , \\
    u_i =0 & \text{in $\Gamma_{\mathrm{b}} \times [0,T]$} \, ,
    
 \label{eq_motions}
\end{cases}
\end{equation}
where $t$ is time, $x_i$ ($i= 1,2,3$) denotes the $i^{th}$ coordinate direction, $x = x_1$, $y = x_2$, and $z = x_3$ denote the streamwise, cross-stream, and vertical coordinate directions, respectively, $u_i$ is the i$^{th}$ filtered velocity component, $\pi$ is a modified filtered pressure field, namely $\pi = \frac{p}{\rho} + \frac{1}{3}\tau_{ii}^{SGS} + \frac{1}{2}u_iu_i $, $\rho$ is a reference constant density, $\tau_{ij}^{SGS}$ is the subgrid-scale (SGS) tensor, $ \Pi_i = \frac{1}{\rho} \frac{\partial p_{\infty}}{\partial x_1} \delta_{i1} + \frac{1}{\rho} \frac{\partial p_{\infty}}{\partial x_2} \delta_{i2}$ is a constant pressure gradient introduced to drive the flow, and $\tilde{f}_i^{\Gamma_{\mathrm{b}}}$ is a forcing term that is used to impose the desired boundary condition at the surface location. $\tilde{f}_i^{\Gamma_{\mathrm{b}}}$ has a finite value at the buildings interface $(\Gamma_{\mathrm{b}})$ and is zero elsewhere. 

The LES algorithm was initially developed in \cite{albertson1999natural, albertson1999surface}. 
Equations are solved in strong form on a regular domain $\Omega$, a pseudo-spectral collocation approach \citep{orszag1969numerical, orszag1970transform} based on truncated Fourier expansions is used in the $x,y$ coordinate directions. 
In contrast, a second-order accurate centered finite differences scheme is adopted in the vertical direction, requiring a staggered grid approach for the $\tilde{u},\tilde{v},\tilde{p}$ variables (these are stored at $(j+1/2)\delta_z$, with $j=1,nz$).
Time integration is performed via a fully explicit second-order accurate Adams-Bashforth scheme, and a fractional step method is adopted to compute the pressure field \citep{chorin1968numerical, kim1985application}.
In addition, nonlinear terms are fully dealiased via the $3/2$ rule to avoid piling up energy in the high wavenumber range \citep{kravchenko1997effect, canuto2007spectral}.
The computational boundary is partitioned as $\partial \Omega = \Gamma_{\mathrm{b}} \cup \Gamma_{\mathrm{t}} \cup \Gamma_{l}$, where $\Gamma_{\mathrm{t}}$ and $\Gamma_{l}$ denote the top and lateral boundaries respectively. 
A free-lid boundary condition applies at $\Gamma_{\mathrm{t}}$, and a no-slip boundary condition is prescribed at $\Gamma_{\mathrm{b}}$ (see equation \ref{eq_motions}). 
An algebraic wall-layer model based on the equilibrium logarithmic law assumption is also applied at $\Gamma_{\mathrm{b}}$ to evaluate tangential SGS stresses at the solid-fluid interface \citep{Chester2007}.
Periodic boundary conditions apply at $\Gamma_{l}$ due to the Fourier spatial representation. 
SGS stresses in the bulk of the flow are parameterized using the \cite{bou2005scale} scale-dependent Lagrangian dynamic Smagorinsky model. 
To model the urban canopy, a discrete forcing approach immersed boundary method is adopted \citep{mohd1997combined, mittal2005immersed, chester2007modeling}.

Over the past two decades, this solver has been used to develop a series of algebraic SGS closure models for the bulk of turbulent flows \citep{meneveau1996lagrangian, porte2000scale, porte2004scale, bou2005scale, lu2010modulated}, wall-layer models \citep{hultmark2013new}, and immersed-boundary methods to accurately represent solid-gas interfaces \citep{tseng2006modeling, chester2007modeling, fang2011towards, li2016impact}. It has also been extensively validated against field and laboratory measurements and used to gain insight into a range of applications involving different flow phenomena, spanning from atmospheric boundary layer flow over flat surfaces to flow over urban areas and forests \citep{bou2005scale, tseng2006modeling,  bou2009effects, fang2011towards, shah2014very, pan2014strong, fang2015large, anderson2015numerical, giometto2016spatial, pan2016estimating, giometto2017direct, giometto2017effects}.

\subsection{High-Fidelity Dataset}
\label{sec:high-fidelity-dataset}
\begin{figure}
\centering
  \includegraphics[width=\textwidth]{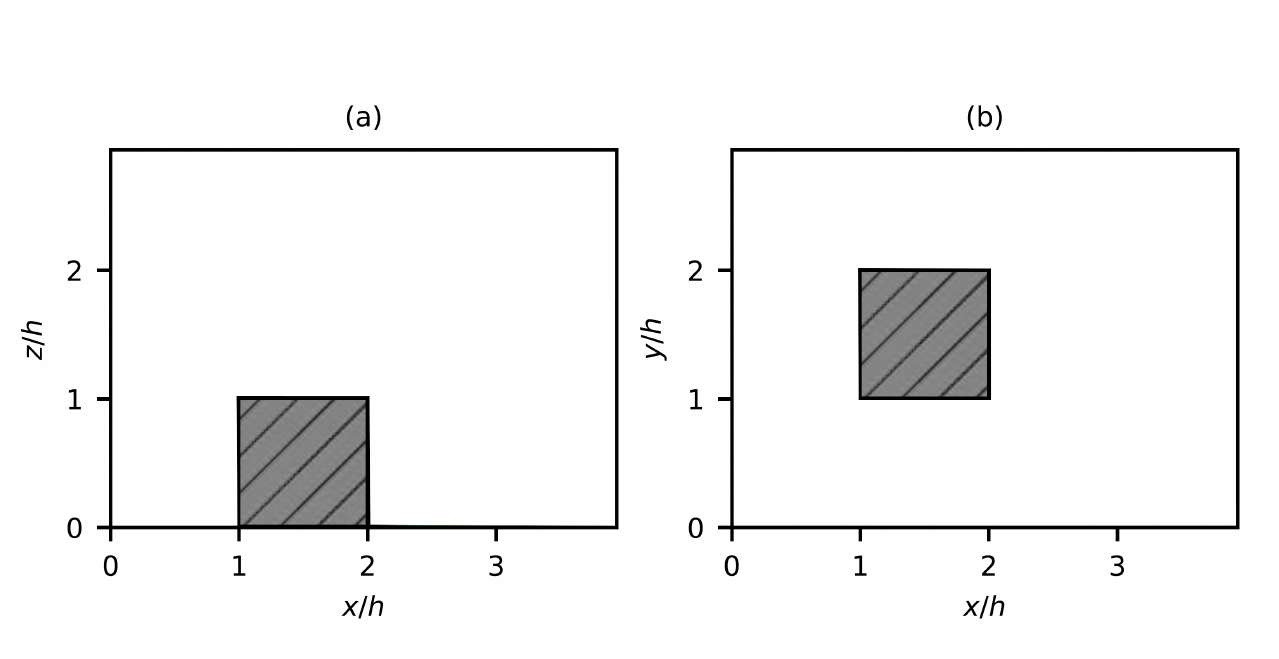}
\caption{Side (a) and planar view (b) of the computational domain.}
\label{config}     
\end{figure}

To generate the training data, a series of LESs of open channel flow over surface-mounted cubes with constant height $h$ are performed, as illustrated in figure \ref{config}.
The computational domain has a size of $\Omega = [0, 4h] \times [0,3h] \times [0,3h]$. 
This is a relatively modest computational domain dimension but will suffice based on the scope of this study. 
The Reynolds number based on friction velocity $u_{\star}$, cube height $h$, and air kinematic viscosity $\nu$ is $Re_{\tau} = 10^{6}$. 
Under these conditions, the flow is in a fully rough aerodynamic regime, and viscous stresses can be safely neglected. 
The aerodynamic roughness length in the inviscid equilibrium wall-layer model at the solid-fluid interface is set to $z_0^{IBM} = 10^{-5} h$.    
The domain is discretized using $N_{x} \times N_{y} \times N_{z} = 64 \times 48 \times 48$ collocation nodes in the streamwise, cross-stream, and vertical directions. 
This results in 16 collocation nodes per cube's edge, which is sufficient to yield resolution-independent results \citep{tseng2006modeling, yang2018numerical}.

Simulations are integrated in time for over $300 T_{e} = 300 (3h)/u_{\star}$, where $T_e$ is the eddy turnover time based on the height of the computational domain and $u_{\star}$. 
Instantaneous velocity snapshots are collected every $0.06*T_e$ during the last $200 T_e$ to evaluate statistically steady-state flow statistics.  
Each simulation took $\approx 30$ hours on 48 cores of Milan CPU @ 2.45GHz compute node; LES model performance will be compared against those from surrogate models in \S \ref{sec:results}.
In this study, we investigate the range of approaching wind angles $\alpha \in [0^{\circ}, 90^{\circ}]$, which encompasses the entire range of flow variability for the considered domain. 
The value of $\alpha$ is controlled via the individual components of the pressure gradient $\Pi_i$. 
The parametric space is uniformly discretized using a step size $\delta \alpha=2.5^{\circ}$ to generate training and test datasets for our models.
We then consider three training datasets: $\alpha_{10}$, $\alpha_{15}$, and $\alpha_{30}$, and a test dataset $\alpha_{test}$.
The $\alpha_{10}$ dataset contains statistics corresponding to $\delta \alpha=10^{\circ}$, i.e., $\alpha=\{0^{\circ}, 10^{\circ}, 20^{\circ}, 30^{\circ}, 40^{\circ}, 50^{\circ}, 60^{\circ}, 70^{\circ}, 80^{\circ}, 90^{\circ}\}$, $\alpha_{15}$ includes statistics corresponding to $\delta \alpha=15^{\circ}$, i.e., $\alpha=\{0^{\circ}, 15^{\circ}, 30^{\circ}, 45^{\circ}, 60^{\circ}, 75^{\circ}, 90^{\circ}\}$, and $\alpha_{30}$ denotes $\alpha$  $=\{0^{\circ}, 30^{\circ}, 60^{\circ}, 90^{\circ}\}$. 
The number of training samples, i.e., input-output pairs in $\alpha_{10}$, $\alpha_{15}$, and $\alpha_{30}$ are 1,474,560, 1,032,192, and 589,824, respectively. 
Moving forward, we will refer to the three training datasets as the big-data ($\alpha_{10}$), moderate-data ($\alpha_{15}$), and small-data ($\alpha_{30}$) regimes.

There is no specific rule of thumb for the number of samples required to train an MLP network, as the success of this task depends on various factors such as the complexity of the problem, the size of the network, the quality of the data, and the desired level of accuracy.
In practice, it is recommended to start with a small number of training samples and gradually increase the dataset's size while monitoring the model's performance on a validation/test set until the desired level of accuracy is achieved \citep{goodfellow2016deep}.
The three training datasets allow us to explore the optimal number of training samples required to train our surrogate models under different data regimes.
To evaluate the performance of the surrogate model on unseen data (test cases), we utilize $\alpha_{test}$. 
The $\alpha_{test}$ comprises of statistics corresponding to $\alpha=$ $\{\alpha \in [0^{\circ}, 90^{\circ}], ~\alpha\equiv (\text{mod} ~ 2.5), ~\alpha\ \notin \{\alpha_{10}, \alpha_{15}, \alpha_{30} \}\}$ and it has 3,538,944 test samples. 
In other words, $\alpha$ spans the range $[0^{\circ}, 90^{\circ}]$, adopting values that are multiples of 2.5 and excludes specific values included in $\alpha_{10}$, $\alpha_{15}$, and $\alpha_{30}$.

In order to accelerate the training process, it is recommended to scale data using pre-processing techniques such as min-max normalization or standardization, typically within the range $[-1, 1]$ or $[0, 1]$ \citep{goodfellow2016deep}. 
In this study, the considered datasets are scaled using standardization techniques that adjust the datasets to have a zero mean and a unit standard deviation.

Throughout the study, $\overline{\theta}$ denotes a time-averaged quantity, $\tilde{\theta}$ denotes a quantity averaged over all of the available approaching wind angles for a given dataset, and $\langle \theta \rangle$ denotes a quantity averaged over horizontal slabs of thickness $\delta_z$ \citep{schmid2019volume}. Based on this convention, a quantity averaged in time and over all approaching wind angles $\alpha$ will be denoted as $\overline{\tilde{\theta}}$.

\begin{figure}
\centering
  \includegraphics[width=\textwidth]{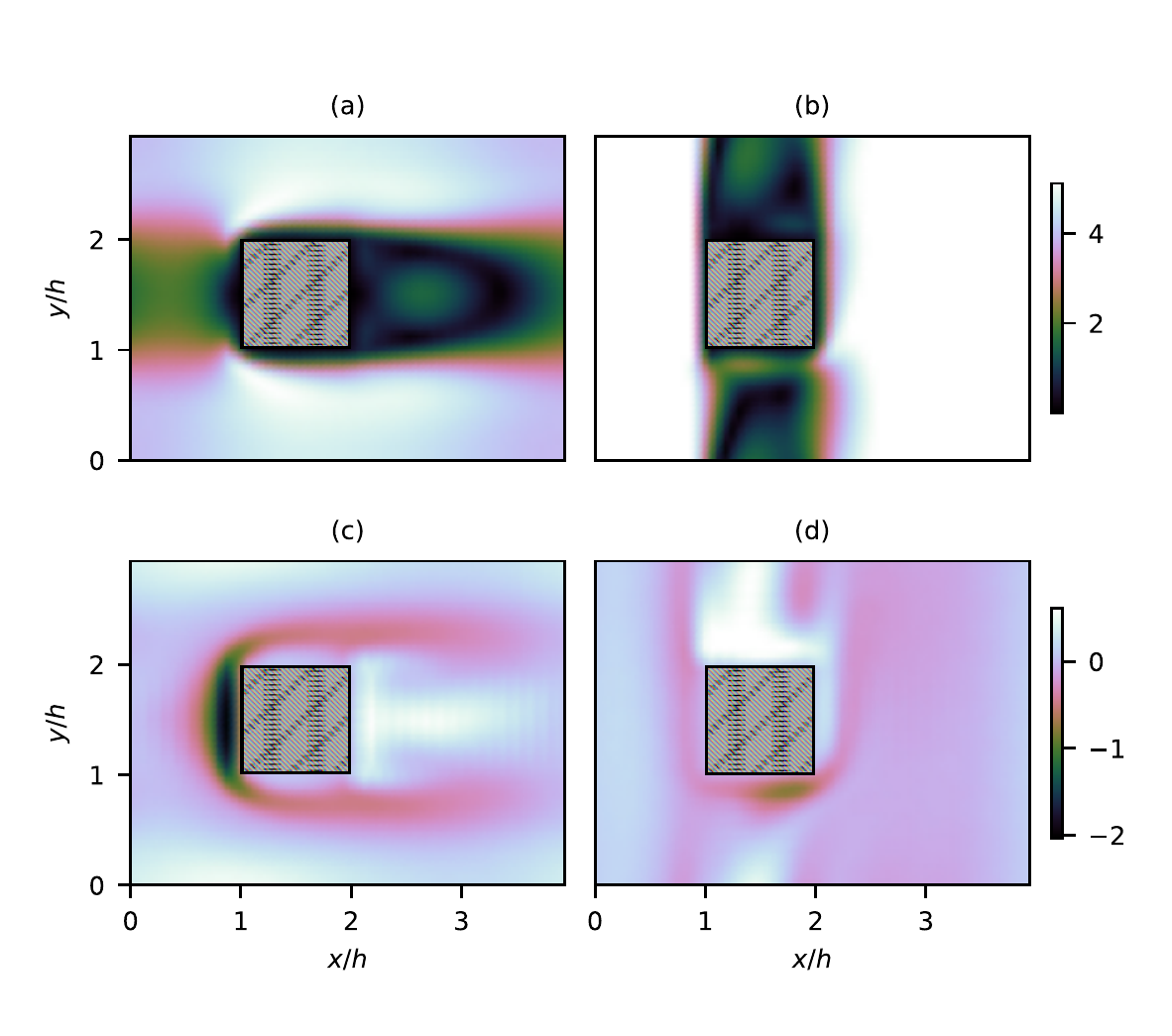}
\caption{Horizontal slice $(x-y)$ of time-averaged horizontal velocity magnitude $\sqrt{\overline{u}^2 + \overline{v}^2}$ (a,b) and time-averaged vertical velocity $\overline{w}$ (c,d) fields at $z/h = 0.5$. Left and right columns correspond to approaching wind angle $\alpha=0^\circ$ and $\alpha=60^\circ$. $\overline{u}$ and $\overline{v}$ are time-averaged streamwise and cross-stream velocity fields, respectively. $z$ denotes the vertical coordinate and $h$ represents the height of the cube.}
\label{fig:mu_u_alpha}     
\end{figure}
To gain a qualitative insight into the spatial variability of the flow fields, figure \ref{fig:mu_u_alpha} shows the spatial variability of the time-averaged horizontal velocity magnitude $(\sqrt{\overline{u}^2 + \overline{v}^2})$ and vertical velocity $(\overline{w})$ fields over a $x-y$ plane at $z/h=0.5$.
Two representative approaching wind angles are considered, namely, $\alpha=0^\circ$ and $\alpha=60^\circ$.

For $\alpha = 0^\circ$, the flow is approximately orthogonal to the cube's face and generates flow statistics that are symmetric about the $y/h=1.5$ axis. Regions of strong flow separation and extensive shear layers characterize the lateral sides of the cube, and high magnitudes of the horizontal time-averaged velocity are observed in the unobstructed flow canyon regions. 
A wake region with weak horizontal velocity and strong positive $\overline{w}$ is also observed in the leeward side of the cube (see figure \ref{fig:mu_u_alpha}, panels $c$ and $d$). 
An apparent horseshoe eddy is also observable from the $\overline{w}$ field, with the head locked in the windward face of the cube and legs extending further downstream.
The $\alpha=60^\circ$ case is characterized by lack of symmetry in the flow and stronger mutual sheltering between cubes. Strong shear layers can still be identified based on the horizontal velocity magnitude plot, and the wake region is significantly skewed. 
Overall, despite the relatively simple canopy geometry, the time-averaged flow field is rich and characterized by coherent patterns that vary significantly with the $\alpha$ coordinate, presenting a challenge for data-driven approaches.

\subsection{Surrogate Modeling}
\label{sec:surrogate-modeling}
This work develops a MLP model mapping the parametric space $\mathbf{X} \in \mathcal{R}^4$ of three-dimensional spatial locations and wind angle to time-averaged flow statistics $\mathbf{Y} \in \mathcal{R}^{9}$. 
The target mapping $\mathcal{M}$ is defined as

\begin{equation}
\mathcal{M}: (\mathbf{x}, \alpha) \xrightarrow[]{} (\overline{u}_i, \overline{u_i^\prime u_j^\prime}) \, , 
\end{equation}
where $\mathbf{x}$ identifies a spatial location in the chosen Cartesian reference system, $\overline{u}_i$ is the $i^\mathrm{th}$ component of the time-averaged velocity at $\mathbf{x}$, and $\overline{u_i^\prime u_j^\prime}$ is the $ij^{th}$ component of the resolved Reynolds stress tensor at $\mathbf{x}$. 
To assess the effectiveness and viability of the proposed model, its predictions are compared against a more traditional alpha-space SPL interpolation technique. 
Details of the proposed MLP and the SPL approaches are discussed in the next sections.

\subsubsection{Multi Layer Perceptron Surrogate}
\label{sec:deep-neural-network}

MLPs are a popular type of Deep Neural Networks (DNNs) and are widely utilized for performing supervised learning tasks such as classification or regression.
They are primarily composed of dense layers.
Each dense layer comprises several perceptrons, also known as neurons, which are densely connected with perceptrons from the preceding layer, i.e., each perceptron in the layer is connected to every perceptron in the previous layer \citep{goodfellow2016deep}.
The transformation performed by a dense layer can be described mathematically as follows: given an input vector $\mathbf{I} \in \mathbf{R}^d$ and a weight matrix $\mathbf{W} \in \mathbf{R}^{l \times d}$, the output $\mathbf{O} \in \mathbf{R}^l$ is computed as:
\begin{equation}
\mathbf{O} = \sigma(\mathbf{WI} + \mathbf{b}) 
\end{equation}
where $\sigma$ is an activation function, $\mathbf{b} \in \mathbf{R}^l$ is a bias vector, and the operator $(+)$ represents element-wise addition. 
The weight matrix $\mathbf{W}$ contains the learnable parameters of the dense layer.
Each row of $\mathbf{W}$ represents the weights connecting a perceptron in the current layer to the perceptrons in the previous layer, and the bias vector $\mathbf{b}$ represents the bias term added to the weighted sum of the inputs.

The choice of $\sigma$ is critical when designing a DNN as it can significantly impact the network's training process and performance on a given task or objective \citep{goodfellow2016deep}.
$\sigma$ introduces nonlinearity into the network, allowing it to learn and represent complex and nonlinear relationships between input and output data.
The sigmoid and hyperbolic tangent (tanh) are commonly used $\sigma$'s but can suffer from the vanishing gradient problem \citep{hochreiter1998vanishing, glorot2010understanding}. 
To address the limitations of the sigmoid and tanh $\sigma$'s, researchers have proposed alternative $\sigma$'s, for instance, the rectified linear units (ReLU) \citep{nair2010rectified}.
However, ReLU is prone to the issue of ``dying ReLU" \citep{xu2015empirical}. 

The Swish $\sigma$ has recently gained attention as a promising alternative to widely-used $\sigma$'s like ReLU and its variants due to its ability to efficiently train DNN while balancing nonlinearity and smoothness effectively \citep{ramachandran2017searching}.
The Swish function is mathematically defined as $f(x) = x \cdot \text{sigmoid}(\beta x)$, where $\text{sigmoid}(\beta x) = [1+\exp(-\beta x)]^{-1}$. 
To evaluate the performance of the Swish function, \cite{ramachandran2017searching} conducted extensive experiments using standard neural network architectures such as ResNet, DenseNet, and Inception, \citep{he2016deep, huang2017densely, szegedy2015going} and datasets including CIFAR-10 and CIFAR-100 \citep{krizhevsky2009learning} and ImageNet \citep{russakovsky2015imagenet}. 
Their results show that the Swish function performs similarly or better than ReLU in terms of accuracy and training efficiency.
Additionally, Swish's non-monotonicity enables greater model expressiveness than ReLU or its variants, allowing it to represent a broader range of functions and patterns, making it an invaluable tool for modeling complex phenomena \citep{ramachandran2017searching}.
Given these properties, the proposed MLP surrogate will be based on the Swish activation function.  

To extract meaningful, complex, and hierarchical representations from the data, we stack multiple dense layers together and refer to them as a DNN \citep{lecun2015deep, goodfellow2016deep}. 
The dense layers between the input and output layers are called hidden layers because they transform the input to extract relevant features for the final output predictions.
MLPs are trained by a backpropagation algorithm proposed in \cite{buscema1998back}. 
Backpropagation computes the gradient of the loss function with respect to trainable parameters using the automatic differentiation technique described in \cite{baydin2018automatic} and iteratively updates weights to minimize the loss. 
The most common choices of the loss function $(\mathcal{L})$ are the mean squared error (MSE) and mean absolute error (MAE) \citep{goodfellow2016deep}. 
These errors can be expressed as
\begin{equation}
\mathcal{L} = \frac{1}{N} \sum_{i}^N ||\mathbf{Y} - \widehat{\mathbf{Y}}||_n \ ,
\end{equation}
with $|| \cdot ||_n$ denoting the $L^n-$ norm, $n=1$ denoting the MAE, $n=2$ the MSE, $\mathbf{Y}$ and $\widehat{\mathbf{Y}}$ the actual and predicted output, and $N$ the number of points in the dataset.

In the context of the $\mathcal{M}$ mapping, we create an MLP-based network of five hidden layers, each with the same number of perceptrons.
As mentioned above, we use Swish non-linearity in all the hidden layers, while the output layer uses a linear activation function. 
To find the optimal architecture, we adopt grid search hyperparameter optimizations scheme \citep{goodfellow2016deep} and only vary the number of perceptrons, considering the set of values $\{32, 64, 128, 256, 512\}$.
Throughout this study, the accuracy of the MLP model is discussed in terms of the relative mean absolute error, i.e., RMAE, defined as 
\begin{equation}
    RMAE = \frac{ \sum  ||\mathbf{Y} - \widehat{\mathbf{Y}} ||_1} { \sum 
 ||\mathbf{Y} ||_1 } \,  , \label{eqn:RMAE}
\end{equation}
%
where $\sum$ denotes summation over points in spatial and/or parametric coordinates. 
RMAE error metrics evenly weigh error magnitudes and provide a measure of the average percentage error in the predictions compared to the ground truth.
The choice of RMAE is also motivated by the use of MAE as a loss function during the training of MLP surrogate models, as will be discussed later in this section.

\begin{table}
  \begin{center}
\def~{\hphantom{0}}
  \begin{tabular}{lccccccccc}
      $\#$ Perceptrons       & 32   & 64   & 128  & 256  & 512 \\[3pt]
      $\#$ Parameters       & 4,681  & 17,545   & 67,849  & 26,6761  & 1,057,801 \\[3pt]
      \textit{$\alpha_{10}$} & 0.091 & 0.070 & 0.058 & 0.059 & 0.058 \\
      \textit{$\alpha_{15}$} & 0.111 & 0.090 & 0.073 & 0.063 & 0.066 \\
      \textit{$\alpha_{30}$} & 0.186 & 0.173 & 0.150 & 0.151 & 0.157 \\
  \end{tabular}
  \caption{ Relative mean absolute error (RMAE) values for different numbers of perceptrons in a five hidden layer multi-layer perceptron (MLP) on test dataset ($\alpha_{test}$). The $\alpha_{10}$, $\alpha_{15}$, and $\alpha_{30}$ are the big-, moderate-, and small-data regime training datasets and $`\#'$ represents the number.}
  \label{NN_Hyp}
  \end{center}
\end{table}
The performance of the resulting networks is evaluated on $\alpha_{test}$ and reported in table \ref{NN_Hyp}. 
The analysis reveals that a shallower network, i.e., a network with five hidden layers and each layer with 32 perceptrons, has the highest RMAE value. 
We can also observe a reduction in the relative errors as we increase the number of perceptrons in each layer, with the performance saturating beyond a certain point.
This phenomenon can be explained using the bias-variance tradeoff, a fundamental concept in ML and statistical modeling \citep{briscoe2011conceptual, goodfellow2016deep}.
\cite{maulik2021turbulent} also observed a similar behavior while using polynomial regression models to develop a mapping between initial conditions, i.e., two-dimensional spatial location and velocity field generated using a low-fidelity numerical algorithm such as a potential solver for Reynolds-averaged Navier-Stokes simulations.

The bias-variance tradeoff describes the relationship between a model's complexity, the accuracy of its predictions, and how well it can make predictions on unseen data that were not used to train the model.
To achieve the right balance between bias and variance, ML methods typically involve selecting a model that is neither too simple nor too complex. 
In the context of MLP, the number of perceptrons determines the model's complexity (see $\#$ Parameters in table \ref{NN_Hyp}), and increasing its complexity helps it learn complex patterns in the data. 
However, increasing the model's complexity also increases the risk of overfitting the training data, which results in a model that does not generalize well to new, unseen data.
Beyond a certain number of perceptron, further increasing the model's complexity does not lead to any significant improvements in performance, resulting in saturation or diminishing returns.

As expected, the performance of DNNs depends on the number of training samples used.
In our experiments, we observe that MLPs trained on a smaller number of training samples, such as $\alpha_{30}$, exhibit lower performance compared to MLPs trained on larger datasets, such as $\alpha_{10}$ and $\alpha_{15}$, regardless of the specific MLP architecture used.
This behavior is consistent with findings from the literature, where it was shown that a larger number of training samples provides a more comprehensive and representative set of information about the underlying distribution of the data \citep{shalev2014understanding, goodfellow2016deep}.
In particular, a larger dataset reduces the risk of overfitting and improves the out-of-sample generalization. 
Moreover, a more extensive dataset size helps reduce the model's bias by providing a more representative sample of the parent data distribution.

\begin{table}
  \begin{center}
\def~{\hphantom{0}}
  \begin{tabular}{lccccccccc}
  Layer & Nodes & Activation  & $\#$ Parameters \\

     Input & 4 & - & - \\ 
    Hidden 1 & 128 & Swish & 640 \\
    Hidden 2 & 128 & Swish & 16512 \\
    Hidden 3 & 128 & Swish & 16512 \\
    Hidden 4 & 128 & Swish & 16512 \\
    Hidden 5 & 128 & Swish & 16512 \\
    Output & 9 & Linear & 1161 \\
  \end{tabular}
  \caption{Architectural details of the MLP-based surrogate model for predicting time-averaged flow statistics ($\overline{u}_i, \overline{u_i^\prime u_j^\prime}$). The $\overline{u}_i$ is the time-averaged mean velocity field in the $i^\mathrm{th}$ direction and $\overline{u_i^\prime u_j^\prime}$ denotes the resolved Reynolds stress tensor.}
  \label{architec}    
  \end{center}
\end{table}
Based on the hyperparameter optimization analysis and bias-variance tradeoff, we adopt an MLP network with five hidden layers containing 128 perceptrons to represent the $\mathcal{M}$ mapping.
This architecture achieves the best performance on our dataset without evidence of overfitting or underfitting. 
The optimal MLP surrogate's architecture is detailed in table \ref{architec}, which has 67,849 trainable and 0 non-trainable parameters.
We acknowledge that more sophisticated hyperparameter tuning methods, such as random search, Bayesian optimization, and search with different network architecture and regularisation schemes recommended in \cite{goodfellow2016deep} could be explored to further improve the model's performance. 

The MLP networks mentioned above are implemented using the TensorFlow machine learning library proposed by \cite{abadi2016tensorflow}. 
 All trainable parameters are initialized randomly using values sampled from a uniform distribution, following the approach of \cite{glorot2010understanding}. 
 The learning rate is set to a constant value of $1\times10^{-3}$ throughout the training process, and the batch size to 1024. The maximum number of epochs is set to 150. 
To introduce non-linear transformations, we relied on a Swish function.
As we worked on a supervised regression task, we adopted MAE as our loss function. 

\subsubsection{Spline Surrogate}
\label{sec:spline}

As mentioned in the opening of this section, the performance of  MLP model is compared against a more traditional SPL surrogate.
The chosen SPL surrogate approximates the $\mathcal{M}_{\mathbf{x}}$ mapping at each spatial location, where 
\begin{equation}
\mathcal{M}_{\mathbf{x}}: \alpha \xrightarrow[]{} (\overline{u}_i, \overline{u_i^\prime u_j^\prime}) \, , 
\end{equation}
whereby $\alpha$ is mapped to each set of statistics $(\overline{u}_i, \overline{u_i^\prime u_j^\prime})$ at the chosen $\mathbf{x}$.
In other words, contrary to the MLP model, the SPL operates without spatial awareness, focusing on interpolating turbulent statistics within the $\alpha$ space only at each $\mathbf{x}$.

Common choices of polynomial functions for splines include linear, quadratic, and cubic spline. 
Cubic splines yield the best performance on $\alpha_{test}$ and are hence chosen for the comparison task. 
The local (in $\alpha$) cubic polynomial for the chosen spline can defined as
\begin{equation}
P_{k}(\alpha) = \beta_{k,0} + \beta_{k,1} (\alpha-\alpha_k) + \beta_{k,2} (\alpha-\alpha_k)^2 + \beta_{k,3} (\alpha-\alpha_k)^3,
\label{eqn:cubic_spline}
\end{equation}
where $\beta_{k,*}$ are the local coefficients for the $k^{th}$ $[\alpha_k, \alpha_{k+1}]$ subinterval, to be determined via enforcement of continuity, smoothness, and adherence to natural boundary conditions at the available data points. 

The RMAE error using cubic SPL surrogate on the $\alpha_{test}$ dataset is 5.0\%, 6.6\%, and 15.8\% in the big-, moderate-, and small-data regimes, respectively. 
We note that the performance evaluation encompasses all nine flow statistics at every spatial location in the computational domain for $\alpha$'s in the $\alpha_{test}$, similarly to what was done when assessing the MLP model.
We also note that the performance of the SPL model based on the considered metrics aligns with that of the MLP model.

The cubic spline is used to predict the flow statistics over a sub-interval $[\alpha_k, \alpha_{k+1}]$, requires $\# =9 \times 4 = 36$ coefficient, where $``9"$ represents both first- and second-order statistics and $``4"$ specifies the number of parameters in the cubic polynomial model. 
To predict flow statistics across big-, moderate-, and small-data regimes, the respective aggregated counts of coefficients for a given spatial location are $\#_{\alpha_{10}}=324$, $\#_{\alpha_{15}}=216$, and $\#_{\alpha_{30}}=108$.
Considering the entire flow system with collocation nodes $N_{x}\times N_{y} \times N_{z}=64 \times 48 \times  48$, the demand for distinct spline models escalates significantly and results in unique $147,456$ models.
Consequently, the total number of parameters required amount to 47,775,744 for $\alpha_{10}$, 31,850,496 for $\alpha_{15}$ , and 15,925,248 for $\alpha_{30}$ data regimes.
Unlike MLP (see \S\ref{sec:deep-neural-network}), which only requires a single model for flow predictions, the SPL-based approach requires thousands of models.
Moreover, the total number of parameters in the SPL model significantly increases, approximately by a factor of $700 \times$ and $235 \times$, in the big- and small-data regimes, respectively. 
This discussion highlights an emerging challenge inherent in the SPL approach, namely, the necessity to manage a large number of distinct models and parameters, which will exacerbate as we increase the resolution of the flow system.

\section{Results}\label{sec:results}
As shown in \S\ref{sec:deep-neural-network} and \S\ref{sec:spline}, both MLP and SPL models yield relatively low RMAE error values for the considered $\alpha_{test}$ dataset. 
Specifically, the MLP (SPL) model achieved an RMAE of 5.8\%, 7.3\%, and 15.0\% (RMAE=5.0\%, 6.6\%, and 15.8\%) on $\alpha_{test}$ when trained using big-, moderate-, and small-data regimes, respectively.
These results imply that the models' ability to capture the $\mathcal{M}$ and $\mathcal{M}_\mathbf{x}$ mapping improves with an increasing amount of training data.
However, while such a metric is frequently used for assessing the performance of data-driven models, it does not provide a direct interpretation of how well the model reproduces individual flow statistics and their spatial variations. 
These quantities are particularly relevant for micrometeorology and urban climate applications. 
Therefore, alternative metrics should be considered to fully assess model performance.

The following sections will address this issue by examining MLP and SPL surrogate models predictions in terms of variable-specific error metrics and spatially-distributed flow statistics. 
Specifically, time-averaged flow statistics from the MLP and SPL models will be compared against corresponding reference LES values for $\alpha$'s within the $\alpha_{test}$ dataset. 
A convergence history analysis is provided in \S\ref{sec:convergence} for the MLP-based surrogate and presents a quantitative analysis of ML-specific error metrics in \S\ref{sec:rel-err-metrics}. 
A physical realizability analysis focusing on mass conservation is discussed in \S\ref{sec:physical-realizability}.
\S\ref{sec:vel-fields} focuses on model performance based on pseudocolor maps of the time-averaged flow field, and \S\ref{sec:mean-flow-profiles} evaluates the predictive ability of the surrogate models in terms of double-averaged turbulent flow profiles for approaching wind angles $12.5^\circ$ and  $35^\circ$.
Finally, in \S\ref{sec:comp-efficiency}, we discuss the computational efficiency of the MLP and SPL surrogates when compared to LES.

\subsection{\label{sec:convergence} Error Convergence for the MLP Model}
\begin{figure}
\centering
  \includegraphics{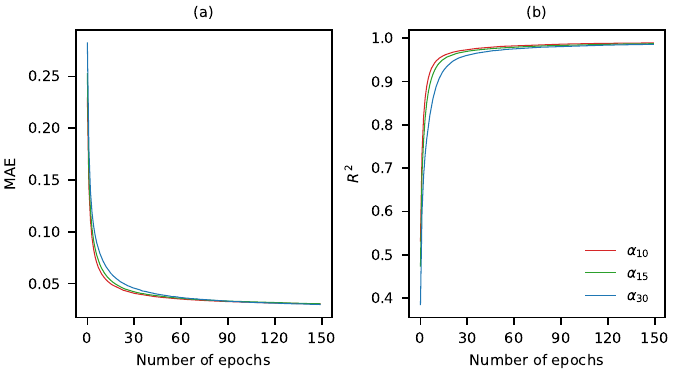}
\caption{Convergence of mean absolute error (MAE) loss (a) and coefficient of determination ($R^2$) (b) for the multi-layer perceptron (MLP)-based surrogate during the training phase. Epochs indicate the number of iterations used in the learning process.  The $\alpha_{10}$, $\alpha_{15}$, and $\alpha_{30}$ are the big-, moderate-,and small-data regime training datasets.}
\label{fig:history}     
\end{figure}

An MLP-based model is trained until convergence is reached, i.e., until when the error between the predicted and actual output values reaches a plateau and the loss function no longer decreases when including additional training epochs. 
At this stage, the weights and biases in the network have been optimized to the point where further iterations no longer lead to significant improvement in performance. 
Figure \ref{fig:history} illustrates the convergence history of the MLP-based surrogate model for a range of pre-defined $\delta \alpha$. 
In addition to MAE, which serves as the loss function for this study, we also utilize the coefficient of determination ($R^2$) as a commonly used metric to evaluate the model performance \citep[see, for e.g.,][]{maulik2021turbulent}. 
Figure \ref{fig:history} shows that the MAE progressively decreases, whereas the $R^2$ improves as the training progresses. 
The error curve plateaus approximately after 150 epochs, with MAE  $\approx 3.0 \times 10^{-2}$, implying that 150 epochs are sufficient to learn the input-output relation. 
Moreover, $R^2 > 0.985$ for all of the training datasets suggests that the MLP  surrogate should accurately capture the desired input-output relation, regardless of the amount of training data.

\subsection{\label{sec:rel-err-metrics}Relative Error Metrics}
To quantify the performance of the MLP model, we compare the relative mean errors for each reconstructed flow statistic against corresponding LES ground truths for $\alpha \in \alpha_{test}$. 
Similar errors are shown for the SPL surrogate.
Given the singular nature of the flow, two distinct regions are examined: the first is the urban canopy layer (UCL), defined as $z/h \in [0, 1]$; the second is above the UCL, i.e., $z/h > 1$.
Flow within the UCL is highly heterogeneous due to the presence of obstacles and is expected to be more challenging for the task at hand when compared to the flow aloft, which is more homogeneous and hence easier to predict.

\begin{table}
  \begin{center}
\def~{\hphantom{0}}
  \begin{tabular}{lccccccccc}
  S.No. & $\overline{u}$ & $\overline{v}$ & $\overline{w}$  & $\overline{u'u'}$  & $\overline{v'v'}$  & $\overline{w'w'}$  & $\overline{u'v'}$  & $\overline{u'w'}$  & $\overline{v'w'}$  \\ 
\textit{$\delta \alpha = 10^{MLP}$} & 10.2 & 11.2 & 21.7 & 7.5 & 8.4 & 6.3 & 16.1 & 20.4 & 21.6 \\ 
\textit{$\delta \alpha = 10^{SPL}$} & 11.6 & 12.8 & 35.0 & 10.6 & 11.6 & 10.5 & 24.7 & 31.7 & 35.5 \\
\textit{$\delta \alpha = 15^{MLP}$} & 14.3 & 11.8 & 24.9 & 9.4 & 10.4 & 8.0 & 18.7 & 23.6 & 24.3 \\
\textit{$\delta \alpha = 15^{SPL}$} & 20.7 & 17.0 & 40.0 & 15.4 & 14.7 & 13.3 & 29.1 & 39.5 & 38.9 \\
\textit{$\delta \alpha = 30^{MLP}$} & 24.5 & 25.8 & 43.4 & 17.1 & 20.5 & 13.7 & 36.5 & 40.6 & 41.6 \\
\textit{$\delta \alpha = 30^{SPL}$} & 52.1 & 64.1 & 57.3 & 30.3 & 51.3 & 19.6 & 53.3 & 57.3 & 64.9 \\
  \end{tabular}
  \caption{Percentage RMAE errors on $\overline{u_i}, \overline{u_i^\prime u_j^\prime}$ across different parametric space discretizations ($\delta \alpha$ sets) using spline (SPL) and MLP surrogate models within the urban canopy layer (UCL), i.e., $z/h \leq 1$. $z$ denotes the vertical coordinate, and $h$ represents the height of the cube.}
  \label{tab:mean_RelL1_in_canopy}    
  \end{center}
\end{table}
Table \ref{tab:mean_RelL1_in_canopy} displays the RMAE values for both first- and second-order flow statistics based on the $\alpha_{test}$ dataset within UCL for the MLP and SPL models.
Based on equation \ref{eqn:RMAE}, this RMAE error corresponds to a summation being performed over all spatial locations within the UCL and $\alpha$.
Interestingly, irrespective of the data regime, both models feature relatively smaller RMAE values for the $\overline{u}$, $\overline{v}$ and normal resolved Reynolds stress components when compared to the RMAE for $\overline{w}$ and resolved shear stress components.
In the big-data regime ($\alpha_{10}$), the RMAE for $\overline{u}$, $\overline{v}$ and normal resolved Reynolds stresses from both surrogates is $<13 \%$, whereas the error increases up to $36\%$ for the $\overline{w}$ component and resolved shear stresses. 
In the moderate- ($\alpha_{15}$) and small- ($\alpha_{30}$) data regimes, the accuracy of both models degrades, with the MLP featuring an overall superior performance.
Specifically, in the $\alpha_{30}$ regime, the MLP prediction features a RMAE of up to $26\%$ for $\overline{u}$, $\overline{v}$, and  resolved normal Reynolds stresses, and $44\%$ for $\overline{w}$ and the resolved shear Reynolds stresses.
Corresponding errors for the SPL surrogate are $64\%$ for $\overline{u}$, $\overline{v}$, and  resolved normal Reynolds stresses, and $65\%$ for $\overline{w}$ and the resolved shear Reynolds stresses. 
Based on these numbers, the performance of the MLP is superior in the UCL across the considered data regimes and features a more gradual decrease than the SPL as the amount of training data is reduced.
\begin{table}
  \begin{center}
\def~{\hphantom{0}}
  \begin{tabular}{lccccccccc}
  S.No. & $\overline{u}$ & $\overline{v}$ & $\overline{w}$  & $\overline{u'u'}$  & $\overline{v'v'}$  & $\overline{w'w'}$  & $\overline{u'v'}$  & $\overline{u'w'}$  & $\overline{v'w'}$  \\ 
\textit{$\delta \alpha = 10^{MLP}$} & 4.5 & 4.1 & 21.7 & 4.8 & 5.2 & 4.3 & 42.4 & 10.5 & 11.5 \\
\textit{$\delta \alpha = 10^{SPL}$} & 4.0 & 3.4 & 25.1 & 5.4 & 5.1 & 5.6 & 44.0 & 13.1 & 14.1 \\
\textit{$\delta \alpha = 15^{MLP}$} & 5.1 & 4.5 & 27.4 & 6.5 & 6.4 & 5.2 & 56.2 & 14.2 & 13.6 \\
\textit{$\delta \alpha = 15^{SPL}$} & 5.4 & 3.8 & 27.8 & 6.9 & 6.2 & 6.2 & 60.0 & 16.4 & 16.8 \\
\textit{$\delta \alpha = 30^{MLP}$} & 13.9 & 16.4 & 46.3 & 10.5 & 12.9 & 8.1 & 92.6 & 23.2 & 25.3 \\
\textit{$\delta \alpha = 30^{SPL}$} & 14.1 & 29.0 & 42.5 & 10.4 & 11.5 & 8.9 & 95.5 & 23.8 & 27.8 \\
  \end{tabular}
  \caption{Percentage RMAE errors on $\overline{u_i}, \overline{u_i^\prime u_j^\prime}$ across different $\delta \alpha$ sets using SPL and MLP surrogate models above the UCL, i.e., $z/h > 1$.}
  \label{tab:mean_RelL1_above_canopy}    
  \end{center}
\end{table}
RMAE errors from the MLP and SPL models in the $z/h > 1$ region are summarized in table \ref{tab:mean_RelL1_above_canopy}.
Both models feature a gradual decrease in performance as the amount of training data is reduced, with overall lower RMAE values across nearly all of the considered flow statistics. 
Notable exceptions include the $\overline{w}$ and $\overline{u^\prime v^\prime}$ statistics, whose behavior will be discussed later.
Discrepancies between the MLP and the SPL model predictions are less apparent than in the UCL, most likely due to the more homogeneous nature of the flow in this region.

\begin{figure}
\centering
  \includegraphics{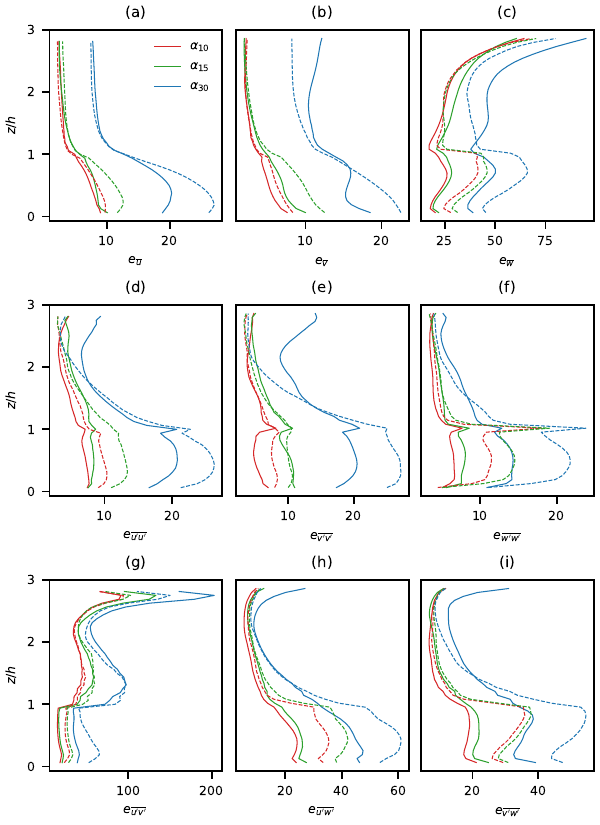}
\caption{Percentage relative mean absolute error (RMAE) $e_i$ for flow statistics averaged in time and space as a function of $z/h$ within $\alpha_{test}$. 
The $e_i$ corresponds to mean velocity fields (panels a, b, c), normal (panels d, e, f), and shear components of resolved Reynolds stresses  (panels g, h, i). The solid and dashed lines represent the MLP and spline (SPL) models, respectively. Color coding is defined in figure \ref{fig:history}.}
\label{fig:RMAE_vert_profiles}     
\end{figure}
To gain more insight into the vertical distribution of model errors, we evaluated RMAE on a horizontal layer-by-layer basis for the MLP and SPL models. 
Based on equation \ref{eqn:RMAE}, this RMAE error corresponds to a summation being performed over horizontal planes and $\alpha$.
To facilitate the discussion, this quantity will be denoted as $e_i$ in the following, where $i$ denotes a given quantity. 
Figure \ref{fig:RMAE_vert_profiles} presents $e_i$ for $\alpha_{10}$, $\alpha_{15}$, and $\alpha_{30}$ using red, green, and blue lines, respectively.
Solid lines denote results obtained using the MLP model, while dashed lines represent those obtained using the SPL model.
The analysis excludes the bottom and top boundaries due to the presence of zero values in certain variables.
Not surprisingly, in the big-data regime, the reconstruction error of both models is generally smaller than in the other two regimes.
Similarly to what was observed in tables \ref{tab:mean_RelL1_in_canopy} and \ref{tab:mean_RelL1_above_canopy}, within the UCL $(z/h \leq 1)$, the MLP model outperforms the SPL, while above the UCL, their performance is comparable.
Larger errors can be observed in the UCL, with peaks in the near-surface region $z=0$ and at the top of the canopy $z = h$.
Notable exceptions to this trend are evident in the $e_{\overline{u^\prime v^\prime}}$ and $e_{\overline{w}}$ statistics, which feature larger errors towards the top of the domain rather than within the UCL.
Upon examination of the values of $\overline{u^\prime v^\prime}$ and $\overline{w}$ near the upper boundary, we observed their magnitudes are relatively small compared to those of other flow statistics;
this small magnitude in the denominator of the corresponding relative error $e_i$ justifies the relatively higher error values observed in figure \ref{fig:RMAE_vert_profiles} and table \ref{tab:mean_RelL1_above_canopy}.
In the moderate-data regime ($\alpha_{15}$), the MLP surrogate outperforms the SPL model for predicting the flow statistics within and above the UCL.
As we transition to small-data regime $(\alpha_{30})$, the relative errors increase for both models across the boundary layer.
Overall, the MLP surrogate demonstrates superior performance within the heterogeneous UCL, while the SPL either outperforms or displays comparable performance to MLP above the UCL.

Quantitatively, in the big-data regime ($\alpha_{10}$), relative errors within the UCL for the MLP vary between a minimum of about 4\% ($e_{\overline{v}}$) to a maximum of $26\%$ ($e_{\overline{ w}}$), while SPL 
features a minimum of $5\%$ ($e_{\overline{w^\prime w^\prime}}$) and a maximum of 42\% ($e_{\overline{w}}$).
Above the UCL, relative errors for both models are below $20\%$, except for $\overline{w}$ and $\overline{u^\prime v^\prime}$.
In the small-data regime ($\alpha_{30}$), relative errors within the UCL for the MLP lies a minimum of 11\% ($e_{\overline{w^\prime w^\prime}}$) to a maximum of $50\%$ ($e_{\overline{w}}$), while SPL features a minimum of $11\%$ ($e_{\overline{w^\prime w^\prime}}$) and a maximum of 66\% ($e_{\overline{w}}$).
Above the UCL, relative errors for both models are below $39\%$, except for $\overline{w}$ and $\overline{u^\prime v^\prime}$.

In general, accurately predicting flow statistics in the UCL is more challenging for the proposed models, with the MLP outperforming the SPL, especially as the amount of training data decreases. 
Larger errors are observed for the $\overline{w}$ and $\overline{u^\prime v^\prime}$ flow statistics throughout.
These findings are consistent with and further support the results presented in tables \ref{tab:mean_RelL1_in_canopy} and \ref{tab:mean_RelL1_above_canopy}, where errors in flow statistics were reported for regions within and above the UCL.
Revisiting the motivation outlined in \S\ref{sec:results}, findings presented in this section highlight that the RMAE values shown in \S\ref{sec:deep-neural-network} and \S\ref{sec:spline} may result in an overestimation of model performance for individual flow statistics, emphasizing the necessity for a more nuanced error analysis in model assessment.

\subsection{\label{sec:physical-realizability} Physical Realizability}
In this section, we evaluate whether the proposed models conserve mass. 
Consistency with such a constraint is especially important in applications involving scalar transport, such as dispersal of seed and plant pathogens spores \citep{chamecki2009large}, pollutant dispersion \citep{britter2003flow}, and snow transport \citep{salesky2019transport}, among others.
Violating such a constraint may lead to spurious oscillation in the concentration field, imbalance in the mass budget, and unphysical negative concentrations \citep{chamecki2008hybrid}.
The residual associated with mass conservation is defined as 
\begin{equation}
    r_c = \iiint_\Omega \frac{h}{u_\star} \frac{\partial \overline{u}_i}{\partial x_i } \delta(\boldsymbol{x}-\boldsymbol{x}^f)d\Omega  \ , 
\end{equation}
where, again, $\Omega$ identifies the computational domain, $\boldsymbol{x}^f$ represents the location of the fluid collocation node, and the Dirac delta function is 1 at fluid nodes and 0 within solid elements. 
\begin{table}
  \begin{center}
\def~{\hphantom{0}}
  \begin{tabular}{lccccccccc}
      $r_c$  &$\mathrm{LES}$ & $\alpha_{10}^{MLP}$ & $\alpha_{15}^{MLP}$ & $\alpha_{30}^{MLP}$ & $\alpha_{10}^{SPL}$  & $\alpha_{15}^{SPL}$   & $\alpha_{30}^{SPL}$ \\[3pt]
     \textit{$\tilde{r}_c$}  & $ 8\times 10^{-8}$ & $ 7\times 10^{-4}$ & $ 1 \times 10^{-3}$  & $4  \times 10^{-3}$ & $ 8 \times 10^{-8}$  & $ 4 \times 10^{-8}$  & $ 8 \times 10^{-8}$ \\ 
     \textit{$\sigma_{r_c}$}& $1 \times 10^{-7}$  & $ 4 \times 10^{-4}$ & $ 7 \times 10^{-4}$  & $3 \times 10^{-3}$ & $ 1 \times 10^{-8}$  & $ 1 \times 10^{-8}$  & $ 1 \times 10^{-8}$ \\
     
  \end{tabular}
  \caption{The $\tilde{r}_c$ and $\sigma_{r_c}$ are the mean and standard deviation of the residual of the mass conservation constraint ($r_c$) values along the $\alpha$ coordinate for the statistics predicted using MLP and SPL surrogate models for various $\alpha$'s in $\alpha_{test}$. }
  \label{table:div}    
  \end{center}
\end{table}
$\tilde{r}_c$ and $\sigma_{r_c}$ for the proposed MLP and SPL models are presented in table \ref{table:div} for $\alpha_{test}$, where $\sigma_{r_c}$ denotes the standard deviation of $r_c$ in $\alpha$ space.
Velocity gradients have been evaluated using discrete Fourier differentiation in the horizontal ($x,y$) directions and second-order accurate centered finite differences in the vertical ($z$), in line with the numerical approach used to evaluate these quantities.
The reference LES flow field satisfies mass conservation with $\tilde{r}_c \approx \sigma_{r_c} \approx 10^{-15}$, i.e., up to machine level precision.
When training the ML model, data are read in single precision, resulting in a target $\tilde{r}_c \approx \sigma_{r_c} \approx 10^{-8}$.
The flow field predicted by the MLP model features $\tilde{r}_c \approx 10^{-3}$ and $\sigma_{r_c} \approx 10^{-3}$ across most of the training sets, which is significantly larger compared to the target LES values. $\sigma_{r_c}$ features a slight deterioration when transitioning to the small-data regime.
The SPL model satisfies the mass conservation constraint to a higher precision, with $\tilde{r}_c$ and $\sigma_{r_c}$ of approximately the same magnitude as the corresponding LES values.
In summary, although the MLP model excels in reconstructing time-averaged mean flow fields, it features relatively larger mass conservation residuals than the SPL model. 
Despite these differences, we note that such residuals would be generally considered as acceptable for most applications involving turbulent transport.

\subsection{\label{sec:flow-field-analysis}Flow Field Analysis}
While the former analyses provide quantitative insight into the quality of the predicted flow field from an RMAE-error perspective across the available $\alpha$ values, they do not provide detailed information about the spatial variability of the reconstructed flow field and associated features (shear layers, wakes, channeling region, etc.).
This section provides a more qualitative albeit fluids-centered discussion of the model's performance in representing the spatial characteristics of the predicted flow field for two representative approaching wind angles, namely $\alpha = 12.5^\circ$ and $\alpha = 35^\circ$. 
These cases encapsulate the observed behavior for the approaching wind angles considered in $\alpha_{test}$, with the $\alpha = 12.5^\circ$ case being representative of flow fields with streamwise direction orthogonal to one of the cube faces and leading to the formation of a stagnation point, and the $\alpha = 35^\circ$ case being representative of flow fields with streamwise direction approaching a lateral cube edge with lack of stagnation points.   
Qualitatively, the proposed surrogate models display similar flow variability in the big- and moderate-data regimes, hence, for brevity, our discussion will be restricted to the big- and small-data regimes only. We will first discuss colormaps of the time-averaged flow field (\S\ref{sec:vel-fields}) and then focus on horizontally- and time-averaged profiles of selected flow statistics (\S\ref{sec:mean-flow-profiles}). 

\subsubsection{\label{sec:vel-fields} Time-Averaged Velocity Fields}
\begin{figure}
\centering
  \includegraphics[width=\textwidth]{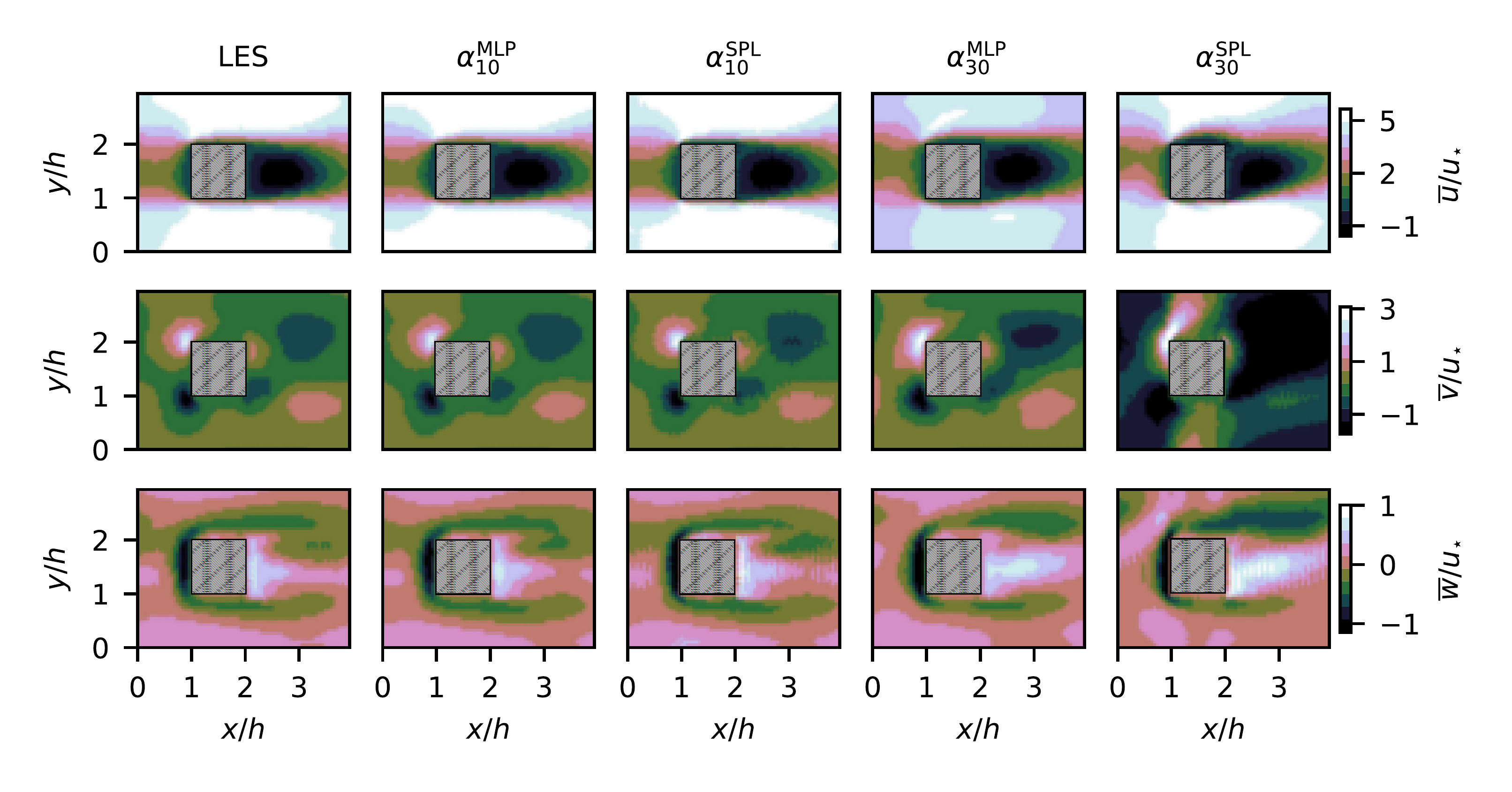}
\caption{Horizontal slice $(x-y)$ of normalized time-averaged streamwise $\overline{u}/u_{\star}$ (top), cross-stream $\overline{v}/u_{\star}$ (middle), and vertical $\overline{w}/u_{\star}$ (bottom) velocity components at $z/h = 0.5$ for $\alpha=12.5^\circ$, where, $u_\star$ represents the friction velocity. Results from the reference large-eddy simulation (LES) are shown in the leftmost column. Predictions using MLP and SPL trained on $\alpha_{10}$ and $\alpha_{30}$ datasets are in the remaining columns. }
\label{fig:recon-fields-12.5}     
\end{figure}

\begin{figure}
\centering
  \includegraphics[width=\textwidth]{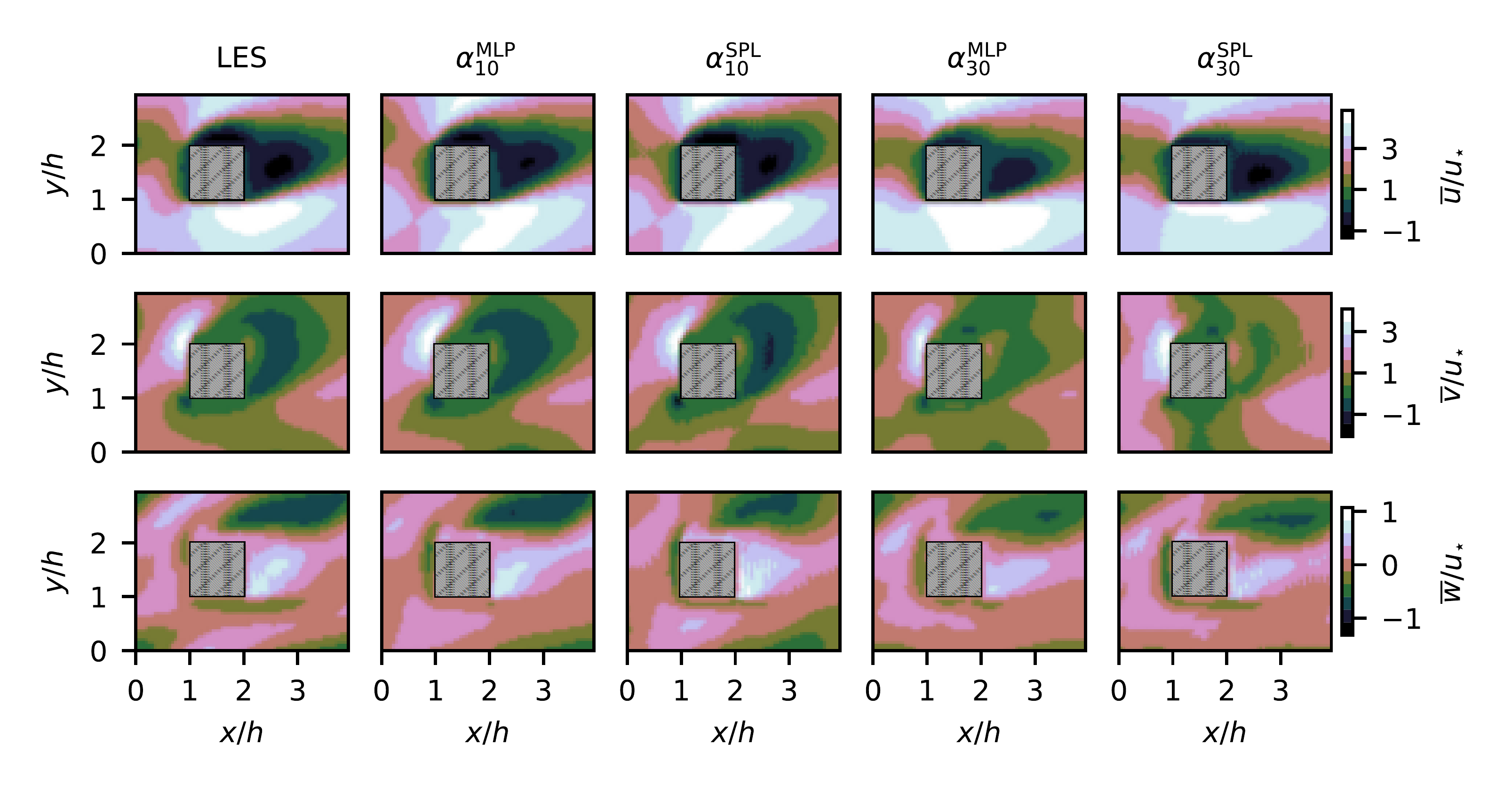}
\caption{Horizontal slice $(x-y)$ of normalized time-averaged velocity components $\overline{u}/u_{\star}$ (top), $\overline{v}/u_{\star}$ (middle), and $\overline{w}/u_{\star}$ (bottom) at $z/h = 0.5$ for $\alpha=35^\circ$. Results from the reference LES are shown in the leftmost column. Predictions using MLP and SPL trained on $\alpha_{10}$ and $\alpha_{30}$ datasets are in the remaining columns.}
\label{fig:recon-fields-35}     
\end{figure}

Figures \ref{fig:recon-fields-12.5} and \ref{fig:recon-fields-35} depict the pseudo-color maps of reference and reconstructed $\overline{u},\overline{v},$ and $\overline{w}$ fields over the $z/h=0.5$ horizontal plane for $\alpha=12.5^\circ$ and $\alpha=35^\circ$, respectively.
The three-dimensional structure of the time-averaged mean flow holds significance as it determines pressure drag and dispersive stresses and largely controls urban ventilation and scalar dispersion in urban environments \citep{coceal2006mean}. 
Hence, it is of interest to characterize model performance in capturing such a quantity. 
For $\alpha = 12.5^\circ$ (figure \ref{fig:recon-fields-12.5}), the reference LES solution features a range of distinct flow features. As the flow approaches the roughness element, $\overline{u}$ decreases under the action of pressure drag and is deflected laterally to accommodate the obstacle (mass conservation). 
At the selected horizontal plane, $\overline{w}$ is negative at the windward side of the cube and remains negative for elongated streams on the lateral sides of the cube. 
This feature signals the presence of a horseshoe vortex with the head locked in place upstream of the roughness element and with counter-rotating legs extending further downstream. 
The flow separates at the lateral sides of the cube and forms two strong shear layers that are especially apparent in the $\overline{u}$ panels. 
The unobstructed flow on the sides of the cubes is characterized by high values of $\overline{u}$. 
On the leeward side of the obstacle, a wake region can be observed, characterized by weak $\overline{u}$ magnitudes, convergence of $\overline{v}$, and positive $\overline{w}$ values. 
Based on the visual inspection, one can conclude that the proposed surrogate models are in very good agreement with the reference LES fields in the big-data regime (see panels corresponding to $\alpha_{10}^\mathrm{MLP}$ and $\alpha_{10}^\mathrm{SPL}$). 
Interestingly, the SPL velocity fields are characterized by minor oscillations, which are especially apparent in the $\overline{w}$ velocity component.
Despite these oscillations, both the MLP and the SPL models appear to be proficient in capturing the salient features of the flow field based on the $\overline{u}$, $\overline{v}$, and also $\overline{w}$ colormaps.
This behavior stands in contrast to findings presented in table \ref{tab:mean_RelL1_in_canopy}, which indicates significant RMAE error within the UCL region for $\overline{w}$ when compared to $\overline{u}$ and $\overline{v}$ components. 
This observation highlights the fact that the RMAE metrics commonly adopted in ML studies alone are not sufficient to characterize surrogate model performance for the considered problem, and physics-based error diagnostics should instead be used to guide the decision-making process, as also mentioned in \cite{wang2020towards}.

In the small-data regime, more noticeable disparities can be observed between the reference solution and the surrogates. 
While both appear to capture the salient features of the $\overline{u}$ and $\overline{w}$ fields, the MLP performs significantly better in terms of $\overline{v}$. 
Both models also tend to overestimate the wake's extent in terms of the $\overline{w}$ velocity, with the MLP offering overall more faithful predictions.
The SPL persists in exhibiting oscillatory artifacts in the flow field, which are again especially apparent in the $\overline{w}$ velocity field.
Overall, the MLP performance appears as qualitatively superior in capturing the intricate nature of the flow field in the small-data regime for the considered plane.

Results for the $\alpha=35^\circ$ approaching angle are shown in figure \ref{fig:recon-fields-35}.
Once again, predictions from both models are in great agreement with corresponding LES values in the big-data regime ($\alpha_{10}$), accurately capturing shear layer regions separating from cube edges and the recirculation region in the wake of the cube (apparent in both the $\overline{v}$ and $\overline{w}$ panels).
The MLP outperforms the SPL both in terms of predicted flow patterns and associated magnitudes. 
Similar to the $\alpha=12.5^\circ$ case, the predictive abilities of both models also decline in the small-data regime $(\alpha_{30})$.
Specifically, both models exhibit apparent discrepancies with the reference LES in terms of the aforementioned flow features.
Unlike in the $\alpha=12.5^\circ$ case, the SPL model performs just as well as the MLP in capturing the $\overline{v}$ field in the small-data regime.
Once again, oscillatory artifacts are observed in the reconstructed $\overline{w}$ when using the SPL model.

Overall, the color maps in figures \ref{fig:recon-fields-12.5} and \ref{fig:recon-fields-35} further demonstrate that MLP and SPL perform relatively well in the big-data regime.
In the small-data regime, the MLP surrogate outperforms the SPL in reconstructing the spatial variability of the flow fields, especially when it comes to predicting flows with approaching wing angles that are approximately orthogonal to the cube's faces.
Findings from this section are consistent with results shown in \S\ref{sec:rel-err-metrics}, where it was concluded that MLP and SPL feature similar RMAE errors in the big-data regime and performance degrades as we transition to the small-data regime, with the MLP overall featuring superior performance as the training data are reduced.
Furthermore, we have also observed that SPL struggles to reconstruct the structure of $\overline{v}$ for flow approaching perpendicular to the cube's faces and consistently introduces oscillatory artifacts across both approaching wind angles for $\overline{w}$.
Although the model predictions for resolved Reynolds stress components are not explicitly shown here, we observe a similar performance difference between the MLP and SPL as the one shown for the mean flow field. 
Specifically, the SPL struggles to predict the spatial structure of $\overline{v^\prime v^\prime}$ and displays significant local errors for $\overline{u^\prime v^\prime}$  and $\overline{v^\prime w^\prime}$ when compared to MLP predictions.

\subsubsection{\label{sec:mean-flow-profiles} Double-Averaged Flow Profiles}
When examining exchange processes between urban areas and the atmosphere, it is common to focus on time- and horizontally-averaged quantities, also known as double-averaged quantities.
Conceptually, the choice of the horizontal averaging region should enable a sensible interpretation of the resulting vertical profiles, retaining the scales of interest while removing scales that will be described statistically \citep{schmid2019volume}. 
In the considered open-channel flow setup, the flow is periodic in the horizontal directions, and the interest is on vertical variations of flow statistics; thus, a sensible spatial-averaging region is a thin slab of thickness $\delta_z$. 
In the past decades, significant attention has been given to developing one-dimensional models for the average wind patterns over urban landscapes that are horizontally homogeneous at the spatial scale of interest \citep{macdonald2000modelling, di2008simple, yang2016exponential, castro2017urban, li2022bridging}.
These profiles, along with associated aerodynamic parameters and profiles of higher-order velocity moments, play a vital role in phenomenological surface-flux models for urban climate and weather forecasting research \citep[see, e.g.,][]{Shamarock2008, Grimmond2010, chen2012research}.
Hence, it is of interest to verify the abilities of the proposed surrogates in reproducing double-averaged profiles of relevant flow statistics. 

\begin{figure}
\centering
  \includegraphics{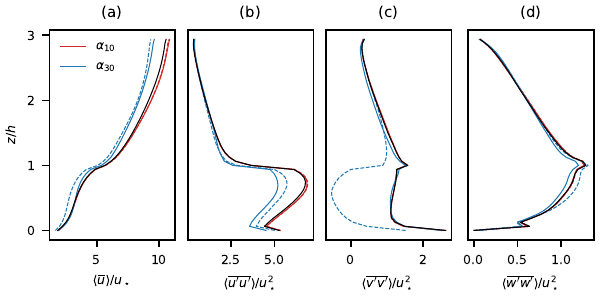}
\caption{Vertical structure of the mean streamwise velocity $\langle \overline{u} \rangle / u_{\star}$ (a), second-order moment of streamwise velocity $\langle \overline{u^\prime u^\prime}\rangle / u_{\star}^2$ (b), cross-streamwise velocity $\langle \overline{v^\prime v^\prime}\rangle / u_{\star}^2$ (c), and vertical velocity $\langle \overline{w^\prime w^\prime}\rangle / u_{\star}^2$ (d) for $\alpha=12.5^\circ$ case. The solid black line in $(a)$, $(b)$, $(c)$, and $(d)$ denotes the reference LES results. The solid and dashed lines represent the MLP and SPL models, respectively.}
\label{fig:stats-12.5}     
\end{figure}

\begin{figure}
\centering
  \includegraphics{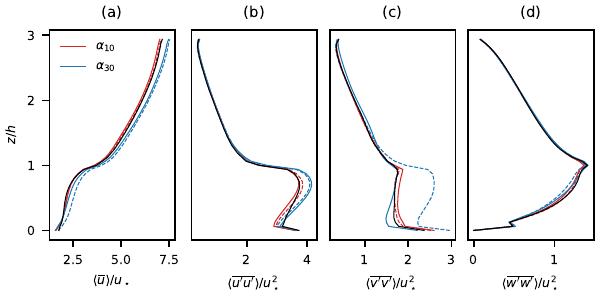}
\caption{Vertical structure of the $\langle \overline{u} \rangle / u_{\star}$ (a), $\langle \overline{u^\prime u^\prime}\rangle / u_{\star}^2$ (b),  $\langle \overline{v^\prime v^\prime}\rangle / u_{\star}^2$ (c),  $\langle \overline{w^\prime w^\prime}\rangle / u_{\star}^2$ (d) for $\alpha=35^\circ$ case. The solid black line in $(a)$, $(b)$, $(c)$, and $(d)$ denotes the reference LES results. The solid and dashed lines represent the MLP and SPL models, respectively.}
\label{fig:stats-35}     
\end{figure}

Figures \ref{fig:stats-12.5} and \ref{fig:stats-35} presents such profiles, with a lens on the mean streamwise velocity ($\langle \overline{u} \rangle$) and normal resolved Reynolds stresses ($\langle \overline{u^\prime u^\prime} \rangle$, $\langle \overline{v^\prime v^\prime} \rangle$,$\langle \overline{w^\prime w^\prime} \rangle$) for the reference approaching wind angles $\alpha=12.5^{\circ}$ and $\alpha=35^{\circ}$, respectively. 
In this analysis, we evaluate profiles in terms of the RMAE error defined as in equation \ref{eqn:RMAE} for within and above UCL regions.
We will first discuss the $\alpha = 12.5^\circ$ profiles.
In the big-data regime, both models perform very well, with profiles pretty much overlapping with those from the reference LES for all considered quantities.
In the small-data regime, the MLP accurately captures $\langle \overline{u}\rangle$ within the UCL, with an RMAE as low as 4\%, whereas the SPL underestimates $\langle \overline{u}\rangle$, with an RMAE of about 15\%. 
Above the UCL, both models underestimate $\langle \overline{u}\rangle$, with an RMAE up to $10\%$.
In the small-data regime, within the UCL, the MLP (SPL) predicts the $\overline{u^\prime u^\prime}$ profile with an RMAE error of 22\% (15\% ), the $\overline{v^\prime v^\prime}$ profile with an RMAE error of 3\% (112\%), and the $\overline{w^\prime w^\prime}$ profile with an RMAE error of 6\% (6\%). 
It is also apparent from figure \ref{fig:stats-12.5} that all models accurately predict normal resolved Reynolds stress above the UCL, with a maximum RMAE error of about 12\%. 
Figure \ref{fig:stats-35} illustrates the vertical profiles of $\langle \overline{u} \rangle$, $\langle \overline{u^\prime u^\prime} \rangle$, $\langle \overline{v^\prime v^\prime} \rangle$, and $\langle \overline{w^\prime w^\prime} \rangle$ corresponding to $\alpha=35^\circ$ case.
In the big-data regime, both models again perform very well, with predicted profiles overlapping with the reference LES solution.  
In the small-data regime, the MLP again accurately captures $\langle \overline{u}\rangle$ within the UCL, with an RMAE as low as 6\%, whereas the SPL overestimates $\langle \overline{u}\rangle$, with an RMAE of 14\%. 
Above the UCL, both models overpredict $\langle \overline{u}\rangle$, with an RMAE up to $6\%$.
Within the UCL, the MLP (SPL) predicts the $\overline{u^\prime u^\prime}$ profile with an RMAE error of 7\% (6\%), the $\overline{v^\prime v^\prime}$ profile with an RMAE error of 5\% (42\%), and the $\overline{w^\prime w^\prime}$ profile with an RMAE error of 4\% (7\%).
Again, all surrogate models show good performance in predicting normal resolved Reynolds stress profiles above the UCL, with RMAE errors below 6\% 

Based on the above analyses, both MLP and SPL surrogates effectively predict double average flow profiles for unseen $\alpha$ values with reasonable accuracy in the big-data regime, capturing the nuanced flow field variability in the considered system. 
As the amount of data decreases, the accuracy of the SPL model drops more significantly than the MLP model, especially within the UCL region, where it is having difficulties capturing the correct velocity magnitudes.  
Despite these discrepancies, both surrogates predict the salient features of mean profiles, including trends in the UCL and above, local maxima and minima, and inflection points.

\subsection{\label{sec:comp-efficiency} Computational Efficiency of the Surrogates}
Surrogate models have been trained on a server featuring an NVIDIA RTX A6000 graphical processing unit (GPUs) and an AMD EPYC 7742 64-Core Processor. 
We report the wall-clock time required for training and inference for both MLP- and SPL-based surrogate models.
The MLP surrogates are trained and run on both the NVIDIA and AMD processors, whereas SPL-based surrogates are evaluated only on AMD processors.
The MLP training time on the NVIDIA cards is 905 s, 608 s, and 310 s for the $\alpha_{10}$, $\alpha_{15}$, and $\alpha_{30}$ training datasets, respectively.
The corresponding training time on the AMD processor is 1670 s, 1100 s, and 615 s, highlighting a solid speedup provided by GPU technology. 
The training time decreases as the size of the training dataset is reduced, as expected, since the number of training samples reduces while the network architecture, batch size, and epochs are held fixed.
As training samples decrease, the model requires fewer iterations to complete an epoch, resulting in a shorter training time.
For the SPL-based surrogate models, the time required to evaluate the polynomial coefficients is 302 s, 300 s, and 294 s for the $\alpha_{10}$, $\alpha_{15}$, and $\alpha_{30}$ training datasets, respectively. 
The inference time for a computational grid of size $64 \times 48 \times 48$ is 6.4 s for the MLP and 4 s for the SPL -- a significant improvement over the 30 hours required to run the LES. 
Specifically, the MLP model speedup is $16{,}875 \times$, whereas the SPL speedup is $27{,}000\times$ across the considered training data regimes.
This analysis reveals that using data-driven surrogates can significantly reduce the computational cost of predicting time-averaged turbulent statistics when compared to LES approaches, thus making these models amenable for use in multi-query approaches such as uncertainty quantification and inverse problems.

\section{Discussion on Model Performance} \label{sec:discussion}
This section offers a critical perspective on the previous findings and diagnoses the cause of observed discrepancies in proposed MLP model predictions. 
The results presented in \S\ref{sec:results} demonstrate that the proposed MLP and baseline SPL surrogates qualitatively capture key features of velocity statistics in the considered flow system, particularly in the big-data regime. 
However, as discussed in \S\ref{sec:rel-err-metrics} and shown in tables \ref{tab:mean_RelL1_in_canopy} and \ref{tab:mean_RelL1_above_canopy}, the accuracy of the proposed models varies depending on the flow variable in question. 
Specifically, the $\overline{u}$ and $\overline{v}$ velocities and normal resolved Reynolds stresses are accurately predicted, whereas predictions of $\overline{w}$ and resolved shear stresses show more significant errors.
Results also show that the accuracy of both MLP and SPL models decreases as the amount of training data is reduced, with the SPL model exhibiting a steep drop in performance between the big- and small-data regimes.

Further, the SPL model demonstrates superior performance compared to the MLP model when it comes to satisfying mass conservation.
This section will provide further insight into these findings and discuss approaches to improve the performance of the proposed MLP model.

%
\begin{figure}
\centering
  \includegraphics{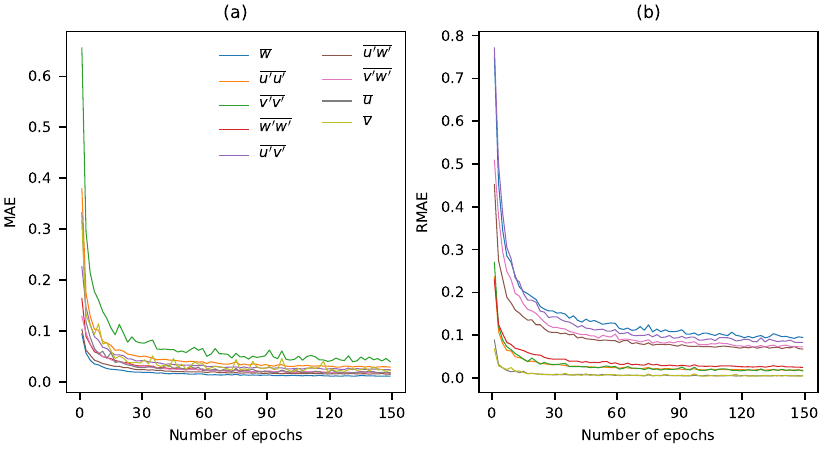}
\caption{Trend of MAE (a) and RMAE (b) for individual flow statistics ($\overline{u_i}, \overline{u_i^\prime u_j^\prime}$) against number of epochs during the training of MLP surrogate model using $\alpha_{10}$ dataset.}
\label{fig:REL_LOSS}     
\end{figure}
%
As mentioned, the MLP model demonstrates variable accuracy in predicting flow statistics.
This behavior can be attributed to the use of a cumulative loss function during the training phase, which does not impose specific constraints on individual flow quantities.
The cumulative loss function, denoted as $L_\mathrm{MLP}$, is defined as the sum of individual loss functions $L_{Y_i}$ for each flow statistic $Y_i$, i.e.,
\begin{equation}
    L_\mathrm{MLP} = \sum_{i=1}^{9}{L_{Y_i}} \, .
\end{equation}
However, evaluating $L_\mathrm{MLP}$ in this manner leads to loss contributions that vary in magnitude, despite the input-standardization procedure discussed in \S\ref{sec:high-fidelity-dataset}. 
As a result, variables with more significant loss contributions are typically prioritized during the optimization process, resulting in a lower RMAE compared to variables with a lesser loss contribution.
This issue is exemplified in figure \ref{fig:REL_LOSS}, where we show the trend of individual loss contributions (MAE) and corresponding RMAE errors during the training phase for the $\alpha_{10}$ dataset. 
From figure \ref{fig:REL_LOSS} $(a)$, it is apparent that quantities such as $\overline{u^\prime u^\prime}$ and $\overline{v^\prime v^\prime}$ (green and orange lines, respectively) feature a larger MAE when compared to, e.g., $\overline{w}$ and $\overline{u^\prime w^\prime}$ (blue and brown lines, respectively). 
Consequently, $\overline{u^\prime u^\prime}$ and $\overline{v^\prime v^\prime}$ have been prioritized during the training process and are ultimately characterized by smaller RMAE values when compared to  $\overline{w}$ and $\overline{u^\prime w^\prime}$.
This explains the observed behavior.
This finding suggests that individually constraining flow statistics in the loss function, e.g., relying on weighted averaged loss contributions, may be a viable pathway to homogenize RMAE errors.
Moreover, the analysis emphasizes the importance of carefully designing the loss function to ensure that the models can effectively capture the relevant information, which can be an extension of this work.

In \S\ref{sec:deep-neural-network} and \S\ref{sec:results}, we have observed a consistent pattern in the performance of the MLP surrogate model depending on the number of training samples used.
MLPs trained on a smaller number of training samples, such as $\alpha_{30}$, exhibit lower performance compared to MLPs trained on larger datasets, such as $\alpha_{10}$ and $\alpha_{15}$.
This behavior aligns with existing literature, where it has established that a larger number of training samples offers a more comprehensive and representative set of information about the underlying distribution of the data \citep{shalev2014understanding, goodfellow2016deep}.
A larger dataset mitigates the risk of overfitting and enhances the out-of-sample generalization. 
Moreover, a more extensive dataset size helps reduce the model's bias by providing a more representative sample of the parent data distribution.

In \S\ref{sec:physical-realizability}, the findings indicate a significant limitation in the performance of the MLP surrogate model regarding its adherence to mass conservation principles.
The proposed MLP network is exclusively trained using observational data, lacking any prior knowledge concerning the underlying physics of the problem.
Recently, it has been shown that incorporating physics into ML formulations can yield improved accuracy, faster training, and improved generalization \citep{karniadakis2021physics}.
Although not explicitly demonstrated in this study, we test the benefits of incorporating physics constraints into the MLP model. 
To achieve this objective, we implemented a physics-informed neural network (PINN) version of our mapping \citep{raissi2019physics}, which enforces mass conservation alongside the observational data.
Results indicate that the inclusion of mass conservation as a soft constraint does not appear to yield any discernible advantages in predicting first-order statistics.
Based on these findings, we can conclude that for the considered flow system, the inclusion of the governing equation, namely mass conservation, as a soft constraint does not yield significant advantages over the proposed MLP surrogate. 
To overcome this limitation, enforcing governing equations through hard constraints might allow the MLP surrogate to precisely satisfy the laws of physics, as suggested in \cite{beucler2021enforcing}.

\section{Conclusion}
\label{sec:conclusions}
In this study, we proposed an MLP-based surrogate model for predicting flow statistics with arbitrary approaching wind angles over an array of surface-mounted cubes. 
The MLP model aimed to approximate the mapping
\begin{equation}
\mathcal{M}: (\mathbf{x}, \alpha) \xrightarrow[]{} (\overline{u}_i, \overline{u_i^\prime u_j^\prime}) \,. 
\end{equation}
MLP model predictions was contrasted against a more traditional method, namely, interpolation in the alpha-space using SPL at every spatial location.
A uniformly spaced grid was adopted to discretize the parameter space of wind angles, i.e., $\alpha$, and three distinct training sets were designed, namely the big- ($\alpha_{10}$), moderate- ($\alpha_{15}$), and small-data ($\alpha_{30}$) regimes. 
Flow statistics ($\overline{u}_i$ and $\overline{u_i^\prime u_j^\prime}$) were generated for each approaching wind angle via LES for training and evaluating the surrogate models.
The trained MLP (SPL) model achieved a global RMAE of 5.8\%, 7.3\%, and 15.0\% (RMAE=5.0\%, 6.6\% and 15.8\%) in the $\alpha_{10}$, $\alpha_{15}$ and $\alpha_{30}$ regime, respectively, when evaluated against the $\alpha_{test}$ dataset. 
To gain insight on the predictive ability of the model beyond the classical RMAE global error used in ML communities, we further examined predictions in terms of RMAE for flow statistics within and above the UCL, and of RMAE a function of height. 
We also discussed the adherence of models to the mass conservation constraint, provided a qualitative examination of predicted time-averaged velocity fields, and analyzed double-averaged profiles.
Key findings from this analysis are summarized in the following.

Within the UCL, the performance of the MLP model was found to be superior across the considered data regimes, featuring a more modest increase in RMAE error than the SPL as the amount of training data is reduced.
Specifically, the average RMAE for MLP mean flow and resolved Reynolds stresses was found to be 14\%, 16\%, and 29\% in the big-, moderate-, and small-data regimes.
Corresponding RMAE values for the SPL model are 20\%, 25\%, and 50\%.
The largest RMAE errors for the MLP model were observed in the $\overline{w}$ and shear stress components, with values ranging from 22\% (big-data) to 44\% (small-data). 
The SPL featured similar trends, with corresponding errors spanning from 36\% to 65\%.  
Trends in the RMAE profiles within the UCL, colormaps of the time-averaged velocity fields, and double-averaged profiles of selected flow statistics further supported these findings.
Both the MLP and SPL models accurately predicted the three-dimensional spatial variability of the velocity fields and double-averaged profiles of mean flow and resolved Reynolds stresses in the big- and moderate-data regime. 
They also effectively captured salient flow features, including shear layers, recirculation regions, flow canyoning, and a host of additional coherent flow patterns. 
More apparent discrepancies instead characterized the small-data regime, with the MLP model generally outperforming the SPL.   
In this regime, both models captured basic modes of variability of the three-dimensional and double-averaged flow statistics, with the SPL exhibiting larger errors, especially for the $\overline{v^\prime v^\prime}$ component.

Above the UCL, the flow is more homogeneous, and both models performed relatively well, featuring a modest and gradual decrease in performance as the amount of training data was reduced.
Specifically, the average RMAE error for mean flow and resolved Reynolds stresses was found to be approximately 13\%, 17\%, and 29\% in the big-, moderate-, and small-data regimes for both models.
The largest errors were observed for the $\overline{w}$ and $\overline{u^\prime v^\prime}$ profiles, with RMAE errors up to 25\% and 44\% in the big-data regime, respectively, and 46\% and 96\% in the small-data one.
These errors can be attributed to the relatively small magnitude of these quantities towards the top of the domain, which yielded relatively larger RMAE values. 

In terms of mass conservation, the MLP exhibited small residual in an absolute sense, albeit significantly larger than those of the reference LES solution and SPL model (MLP residual was $10^{-4}$, residual of reference LES solution was $10^{-8}$).
To address this limitation, we explored enforcing a divergence free condition as soft constraint.
Unfortunately, this effort did not yield notable improvements, highlighting the need for alternative approaches, potentially involving the enforcement of mass conservation as a hard constraint.  

This analysis made clear that while the RMAE error metric is widely accepted for assessing prediction accuracy, it lacks information on the spatial structure and realizability of the flow field, and it also tends to overestimate the error when the target quantity is small in value.
For instance, the RMAE analysis within the UCL showed larger RMAE error values for $\overline{w}$ compared to $\overline{u}$ and $\overline{v}$, but the colormap analysis showed that the structure of all quantities was actually well represented.
This observation points to the necessity of considering additional metrics beyond RMAE to comprehensively evaluate the performance of data-driven models from a physical perspective.

Additionally, the analysis delved into the limitations of the proposed MLP model, specifically addressing its varying accuracy when predicting flow statistics.
This behavior was traced back to the utilization of a cumulative loss function, which averages the MAE loss across all flow statistics without constraining the individual quantity.
Consequently, during the training phase, the model prioritized optimizing quantities with larger loss contributions over those with smaller ones.
This resulted in varying performance in predicting individual statistics during the inference phase.
A potential solution is to use a weighted loss function, which assigns higher importance to flow quantities with smaller loss contributions during training, thereby promoting a more uniform performance of the model across all statistics.

When it comes to model management, a total of 147,456 unique SPL models, each comprising 324, 216, and 108 parameters/coefficients across the big-, moderate-, and small-data regimes, were required.
In contrast, only one MLP model with 67,849 parameters was required for the same task.
The MLP hence presents an advantage in this sense, given that only one model is required regardless of the data size. 
In terms of computational efficiency, both surrogate models offer a substantial speedup, with the MLP and SPL models achieving a speed increase of $10^4 \times$, making them attractive approaches for multi-query applications.

Overall, the MLP approach proved capable of effectively learning nonlinear mapping but required specialized computing infrastructure and resources. 
The SPL performed relatively worse in general, especially in the moderate- and small-data regimes, but was relatively straightforward to implement and deploy with limited computational resources. 
Findings suggest that the selection between the MLP and SPL approaches for a given application should depend on specific requirements, available data, and computational resources. 
When an application has abundant data and advanced computational resources, the MLP surrogate may be preferable due to its superior performance. 
Conversely, if an application has limited computational resources, the SPL approach may be more suitable, providing faster computation and reasonable accuracy at a fraction of the cost.  

Our study demonstrates that MLP-based surrogates are effective in creating approximations of flow statistics over idealized urban areas, as shown in our specific case of neutrally stratified equilibrium flow over surface-mounted cubes and arbitrary approaching wind angles. 
This approach can be readily extended to more complex urban environments, including those with more realistic surface morphologies and atmospheric stability. 
However, developing a generalized version of the proposed models hinges on the availability of extensive CFD datasets and represents a major challenge due to the high computational cost of CFD simulations.  
Nevertheless, the benefits of this approach are significant, as it has the potential to drastically reduce the computational cost of CFD and enable faster design iterations required in uncertainty quantification and inverse problems.

\backsection[Funding]{GSH acknowledges support from the Texas Advanced Computing Center via a Frontera Computational Science Fellowship. This material is also based upon work supported by, or in part by, the U. S. Army Research Laboratory and the U. S. Army Research Office under grant number W911NF-22-1-0178. This work also used the Anvil supercomputer at Purdue University through allocation ATM180022 from the Advanced Cyberinfrastructure Coordination Ecosystem: Services \& Support (ACCESS) program, which National Science Foundation supports grants \#2138259, \#2138286, \#2138307, \#2137603, and \#2138296.}

\backsection[Declaration of interests]{The authors report no conflict of interest.}

\backsection[Data availability statement]{Scripts and datasets generated as part of this study are available from the corresponding author on reasonable request.}

\backsection[Author ORCIDs]{\\
G. S. Hora, \href{https://orcid.org/0009-0004-7894-5175}{https://orcid.org/0009-0004-7894-5175}; \\
M. G. Giometto, \href{https://orcid.org/0000-0001-9661-0599}{https://orcid.org/0000-0001-9661-0599}.}

\bibliographystyle{jfm}
\bibliography{reference_library}

\end{document}